\documentclass[a4paper,twocolumn,superscriptaddress,11pt]{quantumarticle}
\pdfoutput=1
\usepackage[utf8]{inputenc}
\usepackage[english]{babel}
\usepackage[T1]{fontenc}
\usepackage{amsmath}
\usepackage{hyperref}

\usepackage{tikz}
\usepackage{lipsum}

\usepackage{titletoc}
\usepackage[page,header]{appendix}

\usepackage{amsmath}
\usepackage{amsfonts}
\usepackage{amssymb}
\usepackage{amstext}
\usepackage{amsthm}
\usepackage{latexsym}
\usepackage{bbm}
\usepackage{dsfont}
\usepackage{graphicx}
\usepackage{textcomp}
\usepackage{color}
\usepackage{anyfontsize}
\usepackage{mathtools}

\theoremstyle{definition}

\newcommand{\be}{\begin{equation}}
\newcommand{\ee}{\end{equation}}
\newcommand{\bea}{\begin{eqnarray}}
\newcommand{\eea}{\end{eqnarray}}

\newcommand{\white}[1]{\textcolor{white}{#1}}

\newcommand{\real}{{\mathbb R}} %real
\newcommand{\ket}[1]{|#1\rangle} %ket
\newcommand{\bra}[1]{\langle#1|} %bra
\DeclareMathOperator{\tr}{tr}
\newcommand{\1}{\mathbbm{1}}
\newcommand{\ii}{\mathbbm{1}}

\def\>{{\rangle}}
\def\<{{\langle}}

\def\lsim{\mathrel{\rlap{\lower4pt\hbox{\hskip1pt$\sim$}}
		\raise1pt\hbox{$<$}}}                % less than or approx. symbol
\def\gsim{\mathrel{\rlap{\lower4pt\hbox{\hskip1pt$\sim$}}
		\raise1pt\hbox{$>$}}}                % greater than or approx. symbol

\mathtoolsset{showonlyrefs}

\begin{document}

\title{Fundamental energy cost for quantum measurement}
\date{\today}
\author{Kais Abdelkhalek}
\affiliation{Institute for Theoretical Physics,~Leibniz Universit\"at Hannover,~Appelstra\ss{}e 2,~30167 Hannover,~Germany}
\email{kais.abdelkhalek@itp.uni-hannover.de}
\author{Yoshifumi Nakata}
\affiliation{Institute for Theoretical Physics,~Leibniz Universit\"at Hannover,~Appelstra\ss{}e 2,~30167 Hannover,~Germany}
\affiliation{Photon Science Center, Graduate School of Engineering,~The University of Tokyo, and Bunkyo-ku,~Tokyo 113-8656, Japan}
%\affiliation{F\'{\i}sica Te\`{o}rica: Informaci\'{o} i Fen\`{o}mens Qu\`{a}ntics,~Universitat Aut\`{o}noma de  Barcelona,~ES-08193 Bellaterra (Barcelona), Spain}
\author{David Reeb}
\affiliation{Institute for Theoretical Physics,~Leibniz Universit\"at Hannover,~Appelstra\ss{}e 2,~30167 Hannover,~Germany}
%\orcid{0000-0003-0290-4698}

\maketitle

\begin{abstract}
Measurements and feedback are essential in the control of any device operating at the quantum scale and exploiting the features of quantum physics. As the number of quantum components grows, it becomes imperative to consider the energetic expense of such elementary operations.
Here, we derive energy requirements for general quantum measurement, extending previous models and obtaining stronger bounds in relevant situations, and then study two important classes of measurements in detail. One is the projective measurement, where we obtain the exact cost rather than a lower bound, and the other is the so-called inefficient measurement, in which we explicitly show that energy extraction is possible. 
As applications, we derive the energy-precision trade-off in quantum Zeno stabilisation schemes and the exact energy expense for quantum error correction. 
Our results constitute fundamental energetic limitations against which to benchmark implementations of future quantum devices as they grow in complexity.
\end{abstract}

\section{Introduction}
\begin{figure}[t] 
		\centering
		\def\svgwidth{0.5\textwidth}
		\begingroup%
		\makeatletter%
		\providecommand\color[2][]{%
			\errmessage{(Inkscape) Color is used for the text in Inkscape, but the package 'color.sty' is not loaded}%
			\renewcommand\color[2][]{}%
		}%
		\providecommand\transparent[1]{%
			\errmessage{(Inkscape) Transparency is used (non-zero) for the text in Inkscape, but the package 'transparent.sty' is not loaded}%
			\renewcommand\transparent[1]{}%
		}%
		\providecommand\rotatebox[2]{#2}%
		\ifx\svgwidth\undefined%
		\setlength{\unitlength}{3119.89780696bp}%
		\ifx\svgscale\undefined%
		\relax%
		\else%
		\setlength{\unitlength}{\unitlength * \real{\svgscale}}%
		\fi%
		\else%
		\setlength{\unitlength}{\svgwidth}%
		\fi%
		\global\let\svgwidth\undefined%
		\global\let\svgscale\undefined%
		\makeatother%
		\begin{picture}(1,0.46977169)%
		\put(0,0){\includegraphics[width=\unitlength]{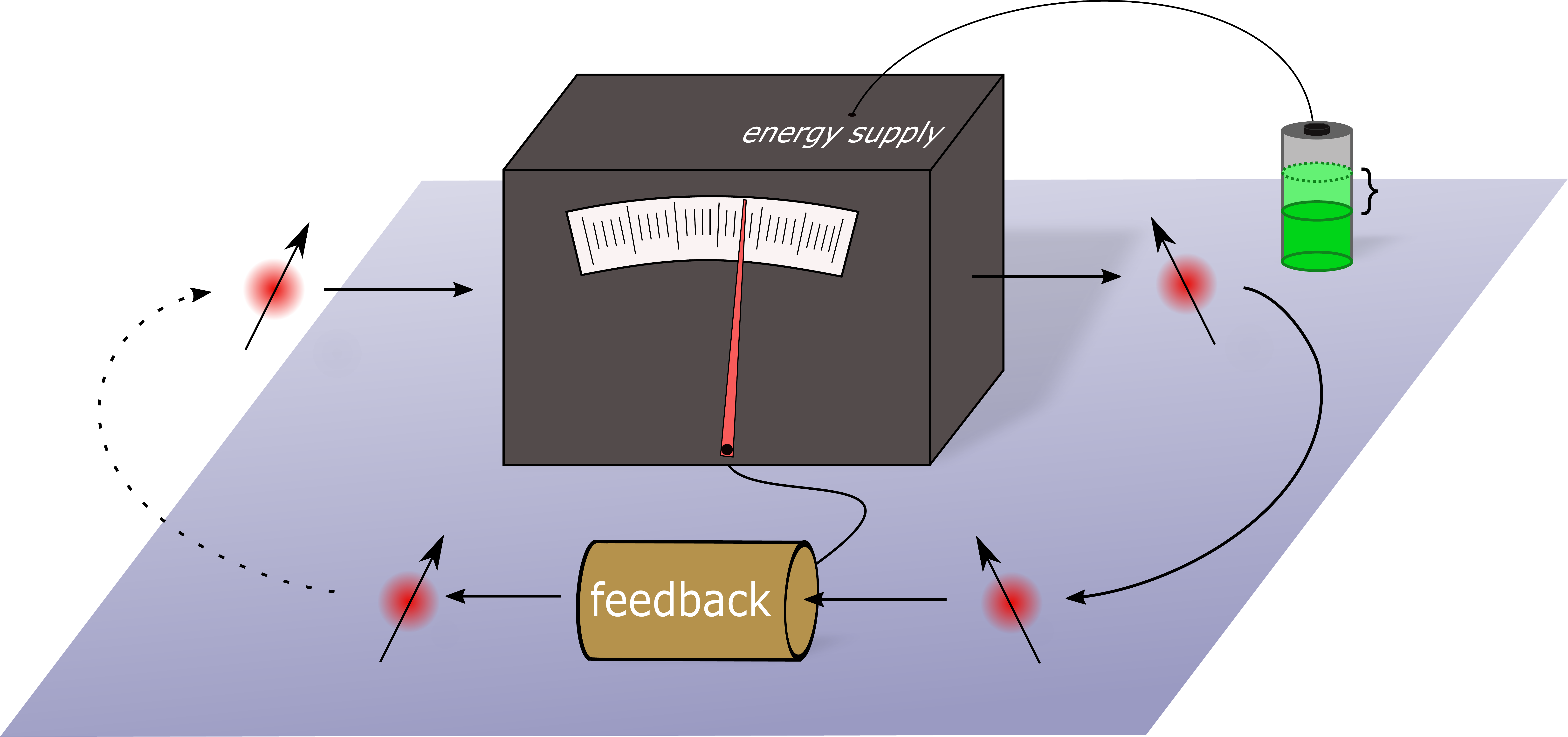}}%
		\put(0.87256584,0.334719368){\color[rgb]{0,0,0}\makebox(0,0)[lb]{\smash{ $E_{\mathrm{cost}}$}}}%
		\end{picture}%
		\endgroup%
		
		\caption{\textbf{Role of quantum measurement.} Most quantum engineering protocols involve frequent measurements to maintain their stability or to control future actions. Whereas measurements are often considered as abstract primitives, we investigate their actual physical implementation and quantify the arising fundamental energy requirements $E_{\rm cost}$.}
		\label{fancy}
	\end{figure}

The ability to manipulate and measure individual quantum systems \cite{haroche-rouchon} enables ever more powerful devices fully exploiting the laws of quantum mechanics,
such as high-precision clocks \cite{wineland-clock}, quantum sensors~\cite{Qmet}, quantum simulators \cite{blatt-q-simulator}, and above all, quantum computers~\cite{Qcomp}.
% as well as the observation of fundamental decoherence processes \cite{haroch-brune}. 
Quantum measurements are crucial for quantum computers to be implemented in a scalable manner~\cite{divincenzocriteria}, for their final readout and more importantly also for continual protection against external noise via error correction \cite{preskill-ftqc}.
Even quantum computation fully based on quantum measurements has been also proposed~\cite{one-way-q-computer,MBQC2,MBQC-experiment}.
%The combination of quantum measurement with feedback is an essential primitive for all these applications, as it allows future actions to depend on past measurement outcomes. 
Thus, with quantum devices becoming increasingly complex, more measurements have to be performed and physical requirements such as the energy supply for these elementary operations must be accounted for (see Fig.\ \ref{fancy}).

This is in parallel to the primitive of information erasure, whose energetic expense will become a limiting technological factor within a few decades \cite{challengesnanoelectronics,mpfrank} as the miniaturization of computers progresses \cite{mooreslaw}. The expense is needed for initializing a computer register and therefore accumulates when using a device repeatedly \cite{bennettreview82}, as is typical in measurement apparatuses. The physical ramifications of information erasure are summarized by Landauer's Principle \cite{landauer}, which demands $k_BT\ln2$ of energy to be dissipated into a heat bath of temperature $T$ for the erasure of each bit of information.
	
So how much energy must be expended for quantum measurements during quantum information processing?
In this paper, based on the purely quantum-mechanical framework similar to the one in Ref.~\cite{sagawaueda08, sagawaueda09}, we first derive the energy cost for general quantum measurements, which goes beyond the usual state-transformation ideas in thermodynamics~\cite{bennettreview82,landauer,kammerlanderanders}.
Compared to many other existing works (e.g. Refs.~\cite{bennettreview82,landauer,sagawaueda08,sagawaueda09,inefficient-measurements,faist-renner}), our framework covers slightly or rather distinct situations and strengthens known results in certain cases.
To illustrate the strength of our form of energy cost, we especially investigate two important classes of quantum measurements, projective and inefficient ones. In the case of projective measurement, the \emph{exact} energy cost is obtained, which is in stark contrast to the previous results about lower bounds on the cost~\cite{sagawaueda08,sagawaueda09}. On the other hand, in the case of inefficient measurements, it turns out that energy can be even extracted during the measurement process if the post-measurement state is irrelevant. This possibility was first suggested in Ref.~\cite{funopaper}, but here, by providing an explicit measurement device for the energy extraction, we prove for the first time that energy extraction is indeed achievable.

Since our measurement model is especially suited when we study the energy cost in quantum information processing, we apply our results to two fundamental processes, quantum Zeno stabilisation~\cite{zeno-original,zeno-lidar} and quantum error correction (QEC)~\cite{divincenzocriteria,preskill-ftqc}, placing fundamental energetic constraints on real-world implementations.
In quantum Zeno control, we derive the precision-energy trade-off relation, which shows that the energy cost should diverge to achieve perfect stabilization.
On the other hand, in QEC, it is revealed that much more energy than previously thought must be expended~\cite{nielsenLandauerQEC}. 
All of these results are not lower bounds of the energy cost but are \emph{exact} costs, and so they will ultimately become the fundamental physical limitation to quantum information processing, in a way similar to the Landauer limit for classical computers~\cite{mpfrank}. 

The paper is organised as follows. We start with describing our measurement model in Sec.~\ref{S:Somm}. Our main results about the energy cost for general quantum measurements are provided in Sec.~\ref{S:ECQM}.  In Sec.~\ref{S:SC}, we study special classes of measurements in detail, projective and inefficient measurements. 
The applications to fundamental protocols in quantum information processing, quantum Zeno stabilisation protocol and error correcting code, are given in Sec.~\ref{S:APPL}. We conclude our paper with summary and discussions in Sec.~\ref{S:sd}.

\section{Setting -our measurement model-} \label{S:Somm}

In this section, we explain the framework in which we study the energy cost of general quantum measurements. The framework is based on the one presented in Ref.~\cite{sagawaueda08, sagawaueda09}.

Unlike the naive expectation that the energy cost is simply given by the energy difference between the states before and after measurement, for counting all energy expense, it is necessary to explicitly consider the physical \emph{implementation} of a measurement device.
The implementation should consist of two steps: the \emph{measurement step} ${\mathcal M}$, to store the measurement outcome in a memory for readout and feedback, and the \emph{resetting step} ${\mathcal R}$, to restore the measurement device to the initial setting so that the device can be used repeatedly.
Taking into account the resetting step is essential because, otherwise, it is possible to `borrow' energy from the measurement device.
The total energy cost of general quantum measurements is obtained only when we consider the costs in these two steps properly.

In the following, we first overview general quantum measurement in Subsec.~\ref{SS:OvM}. The measurement and resetting step are separately explained in Subsecs.~\ref{Sec:MeasM} and~\ref{SS:RS}, respectively.  We then explain how the total energy cost for quantum measurement should be considered in Subsec.~\ref{SS:ToE}.

Before we get started, we here introduce our notation. We denote by $\mathcal{B}(\mathcal{H})$ a set of all bounded operators on the Hilbert space $\mathcal{H}$. Similarly, a set of all quantum states on $\mathcal{H}$ is denoted by $\mathcal{S}(\mathcal{H})$, i.e. $\mathcal{S}(\mathcal{H})= \{\rho \in \mathcal{B}(\mathcal{H}) : \rho \geq 0, \tr [\rho] =1 \}$. To make it clear on which Hilbert space the operator acts on, we often use subscript, e.g. $X_S$ is an operator on $\mathcal{H}_S$. We also denote the marginal states $\tr_M[\rho_{SM}]$ and $\tr_S[\rho_{SM}]$ of $\rho_{SM}$ by $\rho_{S}$ and $\rho_{M}$ for short, respectively.

\subsection{Overview of quantum measurement} \label{SS:OvM}
A quantum measurement on a system $S$ with Hilbert space $\mathcal{H}_S$ is mathematically described by a {\it quantum instrument}, i.e. a set of completely positive maps $\{ T_k \}_{k=1,...,K}$ on $\mathcal{B}(\mathcal{H}_S)$ satisfying $\sum_k T^*_k(\1_S) = \1_S$, where $k$ corresponds to the measurement outcome and $T^*_k$ denotes the adjoint of $T_k$. The action of the map $T_k$ on a state $\rho_S \in \mathcal{S}(\mathcal{H_S})$ can always be written in terms of Kraus operators $M_{ki}$ via $T_k(\rho_S)=\sum_{i=1}^{I(k)} M_{ki} \rho_S M_{ki}^{\dagger}$, where $I(k)$ is often referred to as the Kraus rank of $T_k$. 

A measurement is called {\it efficient} or {\it pure} if $I(k)=1$ for all $k$. 
In addition to this, if $M_k$ are all projection operators, then the measurement is called the \emph{projective} measurement.
A measurement that is not efficient is called {\it inefficient}.
Note that a quantum instrument characterises both the probability $p_k=\tr [T_k(\rho_S)]$ to obtain outcome $k$ and the corresponding post-measurement state $\rho'_{S,k}= T_k(\rho_S) / p_k$.

For a more comprehensive introduction of quantum measurements, we refer to Ref.~\cite{wiseman2009quantum}.

%In contrast, if the post-measurement state can be disregarded and only the outcome probabilities are of interest, it is sufficient to consider a POVM (positive operator valued measure) defined by positive operators $\{E_k\}_{k=1,...,K}$ satisfying $\sum_k E_k =\1_S$. The probability to obtain outcome $k$ is then given by $p_k = \tr [E_k \rho_S]$. Any quantum instrument $\{T_k\}_k$ determines a POVM by $E_k=T_k^{\dagger}(\1_S)=\sum_i M_{ki}^{\dagger} M_{ki}$. 

\subsection{Measurement step ${\mathcal M}$} \label{Sec:MeasM}

When we are interested in the energy cost, it is important to consider all instruments incorporating all relevant systems that are involved in the measurement process.
In particular, the measurement outcome $k$ has to be stored in degrees of freedom of a physical system $M$. 
The register $M$ is often modeled by a quantum system with Hilbert space $\mathcal{H}_M=\bigoplus_{k=1}^K \mathcal{H}_k$ equipped with a Hamiltonian $H_M=\bigoplus_{k=1}^K H_k$, which naturally captures all the important properties one generally demands from a classical memory~\cite{parrondo-nature-physics}. 
The measurement outcome $k$ is stored in a state $\rho'_{M,k} \in \mathcal{S}(\mathcal{H}_M)$, whose support is only on $\mathcal{H}_k$.
In this case, projection operators $\{Q_k\}_{k=1,...,K}$ onto the subspaces $\mathcal{H}_{k}$, respectively, can be applied when one would like to read out the measurement outcome from the register.
Note that the set of projection operators $\{Q_k\}_{k=1,...,K}$ satisfy $\sum_k Q_k =\1_M$.
% with $Q_k^2=Q_k=Q_k^{\dagger}$ for all $k$ and $Q_k\psi_k=\psi_k$ for all $\psi_k\in\mathcal{H}_k$, 

To take into account the register $M$, we define the \emph{measurement step} $\mathcal{M}$ in the following way, which is quite common in the literature~\cite{sagawaueda08,sagawaueda09,parrondo-nature-physics}.
First, a tuple $(\rho_M, U_{SM}, \{Q_k\})$ is called an implementation of a quantum measurement, where $\rho_M$ is the initial state of the memory register $M$, the unitary dynamics $U_{SM}$ describes the interaction between the measured system $S$ and the register $M$, and the projections $\{Q_k\}$ on $M$ with which the outcome $k$ can be read out from the register after measurement.
To any such tuple $(\rho_M, U_{SM}, \{Q_k\})$, a measurement is described by the channel that takes as input an arbitrary initial state $\rho_S$ of $S$ and outputs the post-measurement state
\begin{equation}
\rho'_{SM,k} = (\1 \otimes Q_k) U_{SM} (\rho_S \otimes \rho_M) U_{SM}^{\dagger} (\1 \otimes Q_k) / p_k \label{Eq:zzsig0@}
\end{equation}
on $S$ and $M$ with probability
\begin{equation}
p_k=\tr [(\1 \otimes Q_k) U_{SM} (\rho_S \otimes \rho_M) U_{SM}^{\dagger}] \ ,
\end{equation}
for each $k=1,...,K$. 
We then say that a tuple $(\rho_M, U_{SM}, \{Q_k\})$ is \emph{an implementation of a given measurement $\{ T_k \}_k$} if the associated measurement step outputs the correct post-measurement states on the measured system, $\tr_M[\rho'_{SM,k}] = \rho'_{S,k}= T_k(\rho_S) / p_k$, with correct probability $p_k= \tr [T_k(\rho_S)]$ for all possible input states $\rho_S$. 

The measurement step $\mathcal{M}$ therefore outputs the state
\begin{align}
\rho'_{SM}&=\sum_k p_k \rho'_{SM,k}\\
&=\sum_k (\1 \otimes Q_k) U_{SM} (\rho_S \otimes \rho_M) U_{SM}^{\dagger} (\1 \otimes Q_k), \label{eq:measmodeleeee}
\end{align}
on $S$ and $M$, which correctly stores the outcome $k$ on $M$, since $\rho'_{M,k}\equiv \tr_S [\rho'_{SM,k}]$ has by construction only support on $\mathcal{H}_k$. 
The energy cost for the measurement step $\mathcal{M}$ is defined by
\begin{equation}
\Delta E_{\mathcal{M}} := \Delta E_S + \tr [H_M(\rho'_M - \rho_M)], \label{eq:aekrvdp:aer}
\end{equation}
where $\Delta E_S= \tr [H_S(\rho'_S - \rho_S)]$ with the system Hamiltonian $H_S$ is a trivial energy change in the system due to the measurement. 

Before we go onto the next section, we make two remarks.
First, it can be observed from Eq.~\eqref{eq:measmodeleeee} that the measurement step effectively produces a dephased state, rather than the state fully projected onto the subspace corresponding to some outcome $k$, because the values $k$ are not stored in another register. 
In Appendix~\ref{sec:projectionsokay}, we show that the dephasing operation can be unitarily realised without costing any additional energy, which may be of crucial importance if one would like to link the energy cost, Eq.~\eqref{eq:aekrvdp:aer}, to the thermodynamics work.
Second, for any instrument $\{T_k\}_k$ there exists an implementation $(\rho_M, U_{SM}, \{Q_k\})$. Conversely, any $(\rho_M, U_{SM}, \{Q_k\})$ is an implementation of some instrument $\{T_k\}_k$. 
In this sense, the above measurement model does not place any restrictions and can be considered to be fully general.

\subsection{Resetting step $\mathcal{R}$} \label{SS:RS}
After the measurement, the final state $\rho'_M=\sum_k p_k \rho'_{M,k}$ of the register stores the information of the measurement outcome $k$. This information has to be erased by resetting the register to its initial state $\rho_M$ after the measurement outcome is read out and before the register is used another time. 
We call this process the \emph{resetting step} $\mathcal{R}$, which can be done with the help of a thermal bath $B$ in the following way.

Let $H_B$ be the Hamiltonian of the thermal bath $B$. The state is initially set to be thermal at inverse temperature $\beta=\frac{1}{k_B T}$:
\begin{equation}
\rho'_B=\exp (-\beta H_B)/Z_B
\end{equation} 
where $Z_B=\tr [\exp (-\beta H_B)]$ is the partition function. 
To achieve erasure, the register unitarily interacts with the thermal bath $B$, described by $U_{MB}$, such that its state $\rho'_M$ after $\mathcal{M}$ evolves back to the initial state.
That is, denoting $\rho'_{MB} = \rho'_M \otimes \rho'_B$ and $\rho''_{MB}=U_{MB}(\rho'_M \otimes \rho'_B) U_{MB}^{\dagger}$, the states of $MB$ before and after the resetting step, respectively, the following condition should be satisfied: 
\begin{equation}
\tr_B[U_{MB}(\rho'_M \otimes \rho'_B) U_{MB}^{\dagger}] = \rho_M.
\end{equation} 
This resetting process $\mathcal{R}$ is also known as Landauer erasure \cite{longlandauerpaper}, for which energy cost is naturally defined by
\begin{equation}
\Delta E_{\mathcal R} :={\rm tr}[(H_M+H_B)(\rho''_{MB}-\rho'_{MB})]. \label{eq:erasapp}
\end{equation}

Note that using additional resources, such as a thermal bath $B$, in the resetting step is inevitable since the resetting process generally needs to change the rank and the spectrum of the state, which cannot be done by unitary dynamics only on $M$. In this framework, we consider thermal states a free resource and make use of it because they can easily be obtained by weakly coupling quantum systems to thermal baths at the desired ambient temperature $T$. 
Still, the energy cost of the resetting step, specifically to implement the unitary $U_{MB}$, needs to be accounted for. 

We also emphasise that, in the resetting step ${\mathcal R}$, the measurement outcome $k$ should not be made use of. The whole process in ${\mathcal R}$ must solely rely on a thermal bath $B$ at temperature $T$ as for usual Landauer erasure~\cite{landauer}. This is because the post-measurement state $\rho'_{S,k}$ has usually been altered by feedback following ${\mathcal M}$.
%It is however important that such a physical implementation must yield the correct post-measurement states for \emph{any} input state $\rho_S$.

\subsection{Total energy cost for $\mathcal{M}$ and $\mathcal{R}$}  \label{SS:ToE}

Since the implementation of quantum measurements $\{M_{ki}\}$ consist of the measurement and resetting steps, the overall energy expense is given by the sum of their costs,
\begin{equation}
E_{\rm cost} := \Delta E_{\mathcal M} + \Delta E_{\mathcal R},
\end{equation}
which is the main scope in this paper.

Throughout this paper, we are especially interested in the expression of $E_{\rm cost}$ that is characterised only in terms of system quantities $\rho_S$, $H_S$ and the measurement $\{M_{ki}\}$. This is because, although a quantum measurement $\{M_{ki}\}$ has many different implementations, we are looking for the least expensive in terms of energy. Expressing $E_{\rm cost}$ by the system quantities, we will obtain fundamental results that are independent of the concrete physical measurement implementation, which is in contrast to Ref.~\cite{bennettreview82,esposito-prx}.

	\begin{figure}[t] 
		\centering
		\def \svgwidth{0.6\textwidth}
		\begingroup%
		\makeatletter%
		\providecommand\color[2][]{%
			\errmessage{(Inkscape) Color is used for the text in Inkscape, but the package 'color.sty' is not loaded}%
			\renewcommand\color[2][]{}%
		}%
		\providecommand\transparent[1]{%
			\errmessage{(Inkscape) Transparency is used (non-zero) for the text in Inkscape, but the package 'transparent.sty' is not loaded}%
			\renewcommand\transparent[1]{}%
		}%
		\providecommand\rotatebox[2]{#2}%
		\ifx\svgwidth\undefined%
		\setlength{\unitlength}{838.37534987bp}%
		\ifx\svgscale\undefined%
		\relax%
		\else%
		\setlength{\unitlength}{\unitlength * \real{\svgscale}}%
		\fi%
		\else%
		\setlength{\unitlength}{\svgwidth}%
		\fi%
		\global\let\svgwidth\undefined%
		\global\let\svgscale\undefined%
		\makeatother%
		\begin{picture}(1,0.41051961)%
		\put(0,0){\includegraphics[width=\unitlength,page=1]{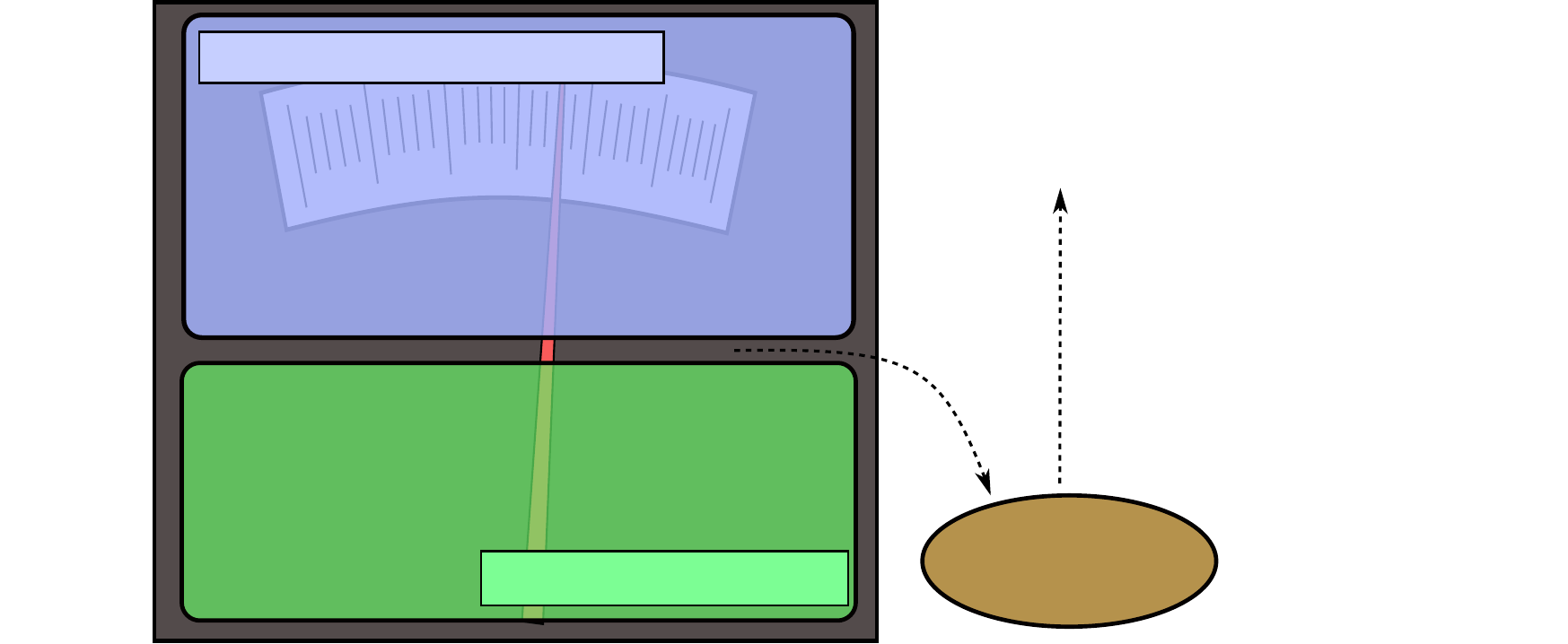}}%
		\put(0.13501263,0.36528133){\color[rgb]{0,0,0}\makebox(0,0)[lb]{\fontsize{8.5pt}{0cm}\selectfont \smash{measurement step $\mathcal{M}$}}}%
		\put(0.31790339,0.03181019){\color[rgb]{0,0,0}\makebox(0,0)[lb]{\fontsize{8.5pt}{0cm}\selectfont \smash{resetting step $\mathcal{R}$}}}%
		\put(0.62957972,0.05672613){\color[rgb]{0,0,0}\makebox(0,0)[lb]{\fontsize{8.5pt}{0cm}\selectfont \smash{external}}}%
		\put(0.6825153,0.18608411){\color[rgb]{0,0,0}\makebox(0,0)[lb]{\smash{$\{V_k\}$}}}%
		\put(0.58593842,0.12719335){\color[rgb]{0,0,0}\makebox(0,0)[lb]{\smash{$k$}}}%
		\put(0.62710068,0.03032527){\color[rgb]{0,0,0}\makebox(0,0)[lb]{\fontsize{8.5pt}{0cm}\selectfont \smash{feedback}}}%
		\put(0,0){\includegraphics[width=\unitlength,page=2]{fig2.pdf}}%
		\put(0.35050894,0.30587445){\color[rgb]{0,0,0}\makebox(0,0)[lb]{\smash{$\1$}}}%
		\put(0.01685217,0.36443782){\color[rgb]{0,0,0}\makebox(0,0)[lb]{\smash{$\rho_{S}$}}}%
		\put(0.14319141,0.23759452){\color[rgb]{0,0,0}\makebox(0,0)[lb]{\smash{$\rho_{M}$}}}%
		%\put(0.57528066,0.36144432){\color[rgb]{0,0,0}\makebox(0,0)[lb]{\smash{$\rho'_{S,k}= \frac{\sum_i M_{ki}\rho_S M_{ki}^{\dagger}}{p_k}$}}}%
		\put(0.67528066,0.36144432){\color[rgb]{0,0,0}\makebox(0,0)[lb]{\smash{$\rho'_{S,k}$}}}%
		\put(0.43524847,0.23705407){\color[rgb]{0,0,0}\makebox(0,0)[lb]{\smash{$\rho'_{M,k}$}}}%
		\put(0.44928924,0.34011055){\color[rgb]{0,0,0}\makebox(0,0)[lb]{\smash{$p_k$}}}%
		\put(0,0){\includegraphics[width=\unitlength,page=3]{fig2.pdf}}%
		\put(0.23342954,0.26856393){\color[rgb]{0,0,0}\makebox(0,0)[lb]{\fontsize{8.5pt}{0cm}\selectfont \smash{$U_{SM}$}}}%
		\put(0.22838398,0.08543517){\color[rgb]{0,0,0}\makebox(0,0)[lb]{\fontsize{8.5pt}{0cm}\selectfont \smash{$U_{MB}$}}}%
		\put(0.34846368,0.08031269){\color[rgb]{0,0,0}\makebox(0,0)[lb]{\fontsize{9pt}{0cm}\selectfont \smash{$\rho_{B}$ thermal}}}%
		\put(0,0){\includegraphics[width=\unitlength,page=4]{fig2.pdf}}%
		\put(0.33959733,0.23445784){\color[rgb]{0,0,0}\makebox(0,0)[lb]{\smash{$Q_k$}}}%
		\put(0.34626699,0.27093354){\color[rgb]{0,0,0}\makebox(0,0)[lb]{\smash{$\otimes$}}}%
		\end{picture}%
		\endgroup%
		
		\caption{{\textbf{Physical implementation of a measurement.}} To perform a general quantum measurement $\{M_{ki}\}$ on state $\rho_S$, system $S$ is input to a measurement device (dark box) that subjects it to a measurement step ${\mathcal M}$ and leaves it in the final state $\rho'_{S,k}$, conditional on the outcome $k$. Subsequent feedback on $S$ is possible as different outcomes belong to orthogonal subspaces on the memory $M$. Before the device is used again, the memory $M$ should be reset to a proper initial state $\rho_M$, using a thermal resource $\rho_B$. This process is called a resetting step ${\mathcal R}$. Hamiltonians $H_S$, $H_M$, $H_B$ determine the total energy cost $E_{\text{cost}}=\Delta E_{\mathcal M}+\Delta E_{\mathcal R}$ required to operate the device physically.
		% The fundamental bounds in \eqref{firstmainresult} and \eqref{secondmainresult} express this cost in terms of system quantities $\rho_S$, $H_S$ and the measurement specification $\{M_{ki}\}$, and thus do \emph{not} depend on microscopic details of the device (blue and green parts).
		}
		\label{schematic}
	\end{figure}

\section{Energy cost of quantum measurement}	\label{S:ECQM}

We now provide our results about the energy cost of general quantum measurement.
Since it can be split into two parts, we derive the energy cost for the measurement and the resetting step in Subsecs.~\ref{SS:ECM} and~\ref{errcostapp}, respectively. The total energy cost is then given in Subsec.~\ref{label-appendix4}.

\subsection{Energy cost for the measurement step $\mathcal{M}$} \label{SS:ECM}
We here show that the energy cost of implementing the measurement step $\mathcal{M}$, defined in Eq.~\eqref{eq:aekrvdp:aer}, splits into a sum of operationally meaningful quantities as follows:
\begin{equation}\label{eq:measeqapp}
\beta \Delta E_{\mathcal{M}} =\beta \Delta E_S + \Delta S+ \beta \Delta F_M + \mathcal{I} + \Delta_Q.
\end{equation}
Here, using Eqs.~\eqref{Eq:zzsig0@},~\eqref{eq:measmodeleeee}, and $\rho_{SM}:=\rho_S\otimes \rho_M$,
each term is given by
\begin{align}
&\Delta E_S= \tr [H_S(\rho'_S - \rho_S)]\\
&\Delta S=  S(\rho_S) - \sum_k p_k S(\rho'_{S,k})\\
&\Delta F_M=F(\rho'_M)-F(\rho_M)\\
&\mathcal{I}= \sum_k p_k I(S:M|k)\\
&\Delta_Q = S(\rho'_{SM})-S(\rho_{SM}).
\end{align}
where $S(\rho)=-\tr [\rho\ln\rho]$ is the von Neumann entropy, $F(\rho)$ is the free energy given by $\tr[\rho H] - S(\rho)/\beta$, and $I(X:Y):=S(\rho_{X})+S(\rho_{Y})- S(\rho_{XY})$ is the a mutual information.
It is clear that each term has an operational meaning that
the $\Delta S$ is the average change of the information about the state,
the $\Delta F_M$ is the purely thermodynamic contribution from the memory in terms of the free energy,
$\mathcal{I}$ is the average amount of correlations built up between $S$ and $M$,
and finally $\Delta_Q$ is the the total entropy increase during the measurement step induced by the projections $\{Q_k\}$. 

Although Eq.~\eqref{eq:measeqapp} may have been implicitly obtained by extending the results in the literature~\cite{sagawaueda08,sagawaueda09}, our derivation is based on less assumptions and generalises the existing work (see Appendix~\ref{label-appendix8} as well). Hence, we explicitly derive Eq.~\eqref{eq:measeqapp} below.

We first note that the post-measurement states $\rho'_{M,k}$ (and hence, also the states $\rho'_{SM,k}$) are mutually orthogonal due to the projection operators $\{Q_k\}$. Denoting the Shannon entropy of the probability distribution $p_k$ by $H(\{p_k\})=-\sum_k p_k \ln p_k$ , the total entropy increase is given by
\begin{widetext}
\begin{align}
\Delta_Q &= S(\rho'_{SM})-S(\rho_{SM})\notag\\
&= H(\{p_k\}) +  \sum_k p_k S(\rho_{SM,k}') - \big( S(\rho_{S}) + S(\rho_{M}) \big)\notag\\
&= H(\{p_k\}) +  \sum_k p_k \left( S(\rho_{S,k}') + S(\rho_{M,k}') - I(S:M|k)\right) - \big( S(\rho_{S}) + S(\rho_{M}) \big)\notag\\
&= S(\rho'_{M}) - S(\rho_{M})  +   \sum_k p_k S(\rho_{S,k}') -  S(\rho_{S}) - \sum_k p_k I(S:M|k) \notag\\
&= S(\rho'_{M}) - S(\rho_{M})  - \Delta S - \mathcal{I} \label{eq:meascostapp1},
\end{align}
\end{widetext}
where we used the additivity of the von Neumann entropy under tensor products, i.e. $S(\rho\otimes \sigma) = S(\rho) + S(\sigma)$ for all states $\rho$ and $\sigma$. 
	
The energy cost of the measurement step is therefore given by
\begin{align}
\beta \Delta E_{\mathcal{M}}&=\beta \Delta E_S + \beta \mathrm{tr} [H_{M}(\rho_{M}' - \rho_{M})]\notag\\
&=\beta \Delta E_S + \beta F(\rho'_M) + S(\rho'_M) \\
&\hspace{27mm} - (\beta F(\rho_M) + S(\rho_M))\notag\\
&=\beta \Delta E_S + \beta \Delta F_M +  \Delta S +  \mathcal{I} + \Delta_Q, \notag
\end{align}	
and we obtain Eq.~\eqref{eq:measeqapp}. $\hfill \blacksquare$

The equation~\eqref{eq:measeqapp} is insightful since it shows how the energy cost for the measurement step $\mathcal{M}$ is split into operationally meaningful quantities. It is however sometimes nicer if the cost is given by quantities that are easy to deal with. To this end, we note that $\mathcal{I}$ and $\Delta_Q$ are non-negative: $\mathcal{I}$ inherits this property from the non-negativity of the mutual information $I(S:M|k)$, whereas $\Delta_Q$ is non-negative because the measurement step $\mathcal{M}$ corresponds to a dephasing channel (see Subsec.~\ref{Sec:MeasM}), which is unital~\cite{nielsenchuang}.
Hence Eq.\ \eqref{eq:measeqapp} immediately implies that
\begin{multline}
\Delta E_{\mathcal{M}} \geq  \Delta E_S \\+  k_B T \Big(S(\rho_S)- \sum_k p_k S(\rho'_{S,k})\Big) + \Delta F_M.
\end{multline}

\subsection{Energy cost for the resetting step $\mathcal{R}$}\label{errcostapp}
Similar to the previous section, we here show that the cost $\Delta E_{\mathcal{R}}$ of the resetting step $\mathcal{R}$ can be expressed as a sum of operational quantities.
Letting $\beta$ be the temperature of a thermal bath, we derive
\begin{equation}\label{eq:erasproof1}
\beta \Delta E_{\mathcal{R}} = - \beta \Delta F_M + \beta \Delta F_B + \mathcal{I}_{MB}.
\end{equation}
Here, $\Delta F_M=F(\rho'_M)-F(\rho_M)$ as defined in the previous section, which also equals to $-(F(\rho''_M)-F(\rho'_M))$, $\Delta F_B:= F(\rho''_B)-F(\rho'_B)$, and $\mathcal{I}_{MB}:=I(M:B)_{\rho''_{MB}}$ is the mutual information built up between $M$ and $B$ during step $\mathcal{R}$.

The derivation of Eq.~\eqref{eq:erasproof1} is straightforward, but we will explicitly derive it for completeness: 
\begin{widetext} 
\begin{align}
\Delta E_{\mathcal{R}}&=\mathrm{tr} [H_{MB}(\rho_{MB}'' - \rho'_{MB})] \notag\\
&= \mathrm{tr} [H_{MB}(\rho_{MB}'' - \rho'_{MB})]  -\frac{1}{\beta} S(\rho_{MB}'') + \frac{1}{\beta} S(\rho'_{MB})\\
&=\mathrm{tr} [H_{MB}(\rho_{MB}'' - \rho'_{MB})] -\frac{1}{\beta} [S(\rho_{M}) +S(\rho_{B}'') -I(M:B)_{\rho''_{MB}} - S(\rho'_{M})-S(\rho'_{B})]\notag\\
&=F(\rho_M) + F(\rho''_B) - F(\rho'_M) - F(\rho'_B) + \frac{1}{\beta}\mathcal{I}_{MB} \notag\\
&= -\Delta F_M + \Delta F_B +\frac{1}{\beta}\mathcal{I}_{MB}, \notag
\end{align}
\end{widetext}
where we used in the second equality that unitary dynamics does not change the entropies, implying that $S(\rho_{MB}'')= S(\rho'_{MB})$. $\hfill \blacksquare$

We further note that $\Delta F_B \geq 0$, which follows from
\begin{align}
\Delta F_B &=\frac{1}{\beta} D(\rho''_B||\rho'_B),
\end{align}
where $D$ is the relative entropy, and Klein's inequality\cite{nielsenchuang}.
It is also the case that $\mathcal{I}_{MB} \geq 0$, which is a basic property of the mutual information.
Hence, it immediately follows from Eq.~\eqref{eq:erasproof1} that
\begin{equation}\label{eq:erascostapp}
\Delta E_{\mathcal{R}} \geq -\Delta F_M.
\end{equation}

%by expressing the free energy in terms of the relative entropy. That is,
%\begin{equation}
%F(\rho)= - \frac{1}{\beta} \ln Z +  \frac{1}{\beta}D(\rho||\rho_{\mathrm{can}}), \notag
%\end{equation}
%where $D(\rho||\sigma):=\tr[\rho(\ln \rho- \ln \sigma)]$ is the relative, and $\rho_{\mathrm{can}}=\mathrm{e}^{-\beta H}/Z$ is a thermal state with the partition function $Z=\tr [\mathrm{e}^{-\beta H}]$.
%It is now immediate to obtain
%\begin{align}
%\Delta F_B &=\frac{1}{\beta} D(\rho''_B||\rho'_B),
%\end{align}
%which is, by Klein's inequality\cite{nielsenchuang}, non-negative.

It has been shown that an optimal process, in the sense that exact equality in the inequality  \eqref{eq:erascostapp} holds, does in general not exist~\cite{longlandauerpaper}. However, if the dimension of the thermal bath $B$ is not restricted, one can approach the lower bound $-\Delta F_M$ arbitrarily closely, e.g. by conducting a process that consists of multiple intermediate steps in which the memory gets temporarily thermalised~\cite{longlandauerpaper,Andersworkdef}. 
Since we are interested in the fundamental bound of the energy cost, we assume in the rest of the paper that the resetting step $\mathcal{R}$ is conducted in such a way that 
\begin{equation}
\Delta E_{\mathcal{R}}= - \Delta F_M.\label{equalityER}
\end{equation}
We however emphasise that this assumption is only for simplification of our results, but does not restrict the validity in a more general setting. Dropping this assumption will only increase the lower bounds by additional non-negative quantities.

\subsection{Total energy cost of a general quantum measurement}\label{label-appendix4}
From Eqs.~\eqref{eq:measeqapp} and~\eqref{equalityER}, we obtain the total energy cost needed to implement a quantum measurement, given by the sum of the cost of step $\mathcal{M}$ and step $\mathcal{R}$,
\begin{equation}\label{eq:overallenergycostlowerbound}
\beta E_{\mathrm{cost}} = \beta \Delta E_S + \Delta S + \mathcal{I} + \Delta. 
\end{equation}
Using the relation $\Delta S + \mathcal{I} + \Delta_Q = S(\rho'_M) - S(\rho_M)$, as shown in the derivation of Eq.\ \eqref{eq:meascostapp1}, we have another expression that
\begin{align}\label{eq:diffent}
\beta E_{\mathrm{cost}} =\beta \Delta E_S + S(\rho'_M) - S(\rho_M),
\end{align}
implying that the total energy cost is essentially determined by the entropy change in the memory.

Although we will make a direct use of Eq.~\eqref{eq:diffent} in later sections, it is important, from the viewpoint of the fundamental limit of the energy cost, to replace the left-hand side with the quantities that sorely depend on the system. 
Using the non-negativity of $\mathcal{I}$ and $\Delta_Q$, we have from Eq.~\eqref{eq:overallenergycostlowerbound} that
\begin{equation} \label{eq:firstmainresult}
E_{\mathrm{cost}} \geq \Delta E_S + k_B T \big[S(\rho_S) - \sum_k p_k S(\rho'_{S,k})\big],
\end{equation}
which is independent of the specific measurement implementation $(U_{SM},\rho_M, \{Q_k\})$.

Note that Eq.~\eqref{eq:diffent} is consistent with applying Landauer's principle~\cite{landauer,longlandauerpaper} to the measurement process, as has been first done by Bennett \cite{bennettreview82} and was generalised to feedback protocols by Ref.~\cite{esposito-prx}. 
We however emphasise that our framework derives Eq.~\eqref{eq:firstmainresult} in the purely quantum mechanical framework. 

Our derivation is also similar to the previous one in Ref.~\cite{sagawaueda08,sagawaueda09,inefficient-measurements}, but extends to more general case because, unlike the previous derivation, ours can be applied to inefficient measurements without assuming thermality of $\rho_S$. This generalization enables us to argue the energy cost of important quantum measurement in a much more concrete manner, as we will see in later sections.

See Appendix~\ref{label-appendix8} for the comparison with previously known results~\cite{bennettreview82,landauer,sagawaueda08,sagawaueda09,inefficient-measurements,faist-renner}, including those not explained here, in more detail.

\section{Special cases} \label{S:SC}

In this section, to illustrate the strength of our results Eqs.~\eqref{eq:firstmainresult} and~\eqref{secondmainresulter}, we consider two special cases of measurements: \emph{projective measurements} in Subsec.~\ref{sec:appprof} and \emph{inefficient measurements} in Subsec.~\ref{sec:appinefficiency}. 

\subsection{Case 1: projective measurements}\label{sec:appprof}
Projective measurements are the textbook examples of ``standard'' quantum measurements. They are described by projective measurement operators $M_k=P_k$ with $P_k^2=P_k^{\dagger}=P_k$ and map the initial state $\rho_S$ of the measured system to the post-measurement state $\rho'_{S,k}=P_k\rho_S P_k/p_k$ with probability $p_k=\tr [P_k\rho_S]$. In particular, projective measurements belong to the class of efficient measurements due to the one-to-one correspondence between measurement operator $P_k$ and outcome $k$. 

A measurement device that implements such a projective measurement $\{P_k\}$ on $S$ is described by a tuple $(U_{SM},\rho_M,\{Q_k\})$ satisfying
\begin{multline}\label{eq:projectionrequirement2}
\tr_M\left[(\1 \otimes Q_k) U_{SM} (\rho_S\otimes \rho_M) U_{SM}^{\dagger} (\1 \otimes Q_k)\right] \\=  P_k \rho_S P_k \quad \forall \rho_S,  \forall k.
\end{multline}
We require this equality to hold for any states $\rho_S$.
%, otherwise the device does not perform the projective measurement on all possible input states.

Due to Eq.~\eqref{eq:diffent}, to obtain the energy cost $E_{\rm proj}$ of projective measurements, it suffices to compute the entropy difference in the memory.
While the entropy difference is in general hard to compute, it can be simplified in the case of projective measurement.
Denoting $\tr_S \big[U_{SM}(\rho_S\otimes \rho_M)U_{SM}^{\dagger}\big]$ by $\tilde{\rho}_M$, the state of the memory after the measurement is given by $\rho'_M=\sum_k Q_k \tilde{\rho}_M Q_k$. 
Using the fact that $(U_{SM},\rho_M,\{Q_k\})$ is, by assumption, an implementation of the projective measurement $\{P_k\}$, we can show that the state $\tilde{\rho}_M$ takes the form
\begin{align}
\tilde{\rho}_M=\sum_k \tr [P_k \rho_S] \sigma_{M,k},
\end{align}
where the $\sigma_{M,k}= Q_k \sigma_{M,k} Q_k$ are mutually orthogonal and have entropy $S(\sigma_{M,k})=S(\rho_M)$ for all $k$ (see Appendix~\ref{App:Lemma}). 
The post-measurement state of the memory is therefore given by
\begin{align}
\rho'_M &=\sum_k Q_k  \Big(\sum_{k'} \tr [P_{k'} \rho_S] \sigma_{M,k'} \Big) Q_{k}\\
&= \sum_k \tr [P_k \rho_S] \sigma_{M,k}\\
&= \sum_k p_k \sigma_{M,k},
\end{align}
implying that $S(\rho'_M) = H(\{p_k\}) + S(\rho_M)$, where $H(\{p_k\})=-\sum_k p_k \ln p_k$ is the Shannon entropy.
Hence, we obtain that
\begin{equation}\label{secondmainresulter}
\beta E_{\mathrm{proj}} = \beta \Delta E_S + H(\{p_k\}).
\end{equation}

This result is a significant improvement over Eq.~\eqref{eq:firstmainresult} unless $\rho_S$ was already classical, in which case both results agree. 
It thus reveals that quantum coherences require much more energy for measurement than previously known. 
The equation~\eqref{secondmainresulter} is strong also from the viewpoint that, similar to single-shot treatments of general processes~\cite{faist-renner}, it is derived directly from quantum mechanics and goes beyond the traditional state-transformation ideas in thermodynamics~\cite{bedingham-maroney}. 
We compare Eq.~\eqref{secondmainresulter} with the result obtained in Ref.~\cite{faist-renner}, which disagrees since the situations are different, in Appendix~\ref{label-appendix8}.

\subsection{Special case 2: inefficient measurements}\label{sec:appinefficiency}

Another interesting class of quantum measurment is an \emph{inefficient} measurement.
As explained in Subsec.~\ref{SS:OvM}, a measurement is said to be inefficient if the measurement changes the state as $\rho_S\mapsto\rho'_S=\sum_{k,i} M_{ki}\rho_S M_{ki}^{\dagger}$. The index $i$ ranges from $1$ to the Kraus rank $I(k)$ of the channel $T_k$.
We henceforth call the maximal Kraus rank of all elements $T_k$ of a given quantum instrument the {\it inefficiency $I$} of the quantum instrument $\{T_k\}$. Clearly, if $I=1$ we recover the case of efficient measurements.

To derive the energy cost in this case, let us denote by $p_k= \tr \big[\sum_{i} M_{ki}\rho_S M_{ki}^{\dagger}\big]$ the probability of receiving outcome $k$ and by $\rho'_{S,k}= \sum_{i} M_{ki}\rho_S M_{ki}^{\dagger} / p_k$ the corresponding post-measurement state on $S$. Furthermore, define $r_{ki}:= \tr \big[M_{ki}\rho_S M_{ki}^{\dagger}\big]$. We then have 
\begin{widetext}
\begin{align}
\sum_k p_k S(\rho_{S,k})
%&=\sum_k p_k S\Big(\sum_i \frac{M_{ki}\rho_S M_{ki}^{\dagger}}{p_k}\Big) \\
&=\sum_k p_k S\Big(\sum_i \frac{r_{ki}}{p_k}\frac{M_{ki}\rho_S M_{ki}^{\dagger}}{r_{ki}}\Big) \\
&\leq \sum_k p_k \left[ H\Big(\Big\{\frac{r_{ki}}{p_k}\Big\}_i\Big) + \sum_i \frac{r_{ki}}{p_k}S\Big(\frac{M_{ki}\rho_S M_{ki}^{\dagger}}{r_{ki}}\Big) \right] \label{eq:ineff1}\\
&\leq \sum_k p_k \left[ \ln I + \sum_i \frac{r_{ki}}{p_k}S\Big(\frac{\sqrt{\rho_S}M_{ki}^{\dagger} M_{ki}\sqrt{\rho_S}}{r_{ki}}\Big) \right] \label{eq:ineff2}\\
%&= \ln I + \sum_{k,i} r_{ki} S\Big(\frac{\sqrt{\rho_S}M_{ki}^{\dagger} M_{ki}\sqrt{\rho_S}}{r_{ki}}\Big) \\
&\leq \ln I + S\Big(\sum_{k,i} \sqrt{\rho_S}M_{ki}^{\dagger} M_{ki}\sqrt{\rho_S}\Big) \label{eq:ineff3}\\
&= \ln I + S(\rho_S),
\end{align}
\end{widetext}
where Ineq.~\eqref{eq:ineff1} and \eqref{eq:ineff3} follow from the well-known property of the von Neumann entropy, Ineq.~~\eqref{eq:ineff2} holds because the Shannon entropy of any probability distribution with $I$ elements is bounded from above by $\ln I$ and $S(L L^{\dagger})=S(L^{\dagger} L)$ for any linear operator $L$.
Together with Ineq.~\eqref{eq:firstmainresult}, we obtain 
\begin{equation}\label{eq:ineffapp}
\beta E_{\mathrm{cost}} \geq \beta \Delta E_S - \ln I.
\end{equation}

Noting that the right-hand side of Ineq.~\eqref{eq:ineffapp} can be negative when $I>1$, it suggests that it may be possible to \emph{extract} a positive amount of energy in the process of quantum measurement. This possibility was first pointed out in Ref.~\cite{funopaper} by similarly showing in a different setting that a lower bound of the energy cost can be negative.
It was however not addressed whether the bound is tight and, if it is, how to achieve the bound.
We below provide an explicit construction of the measurement device that saturates Ineq.~\eqref{eq:ineffapp}, showing the tightness of Ineq.~\eqref{eq:ineffapp} and proving for the first time that energy can be indeed extracted when the measurement is inefficient.
%It should be also noted that Ineq.~\eqref{eq:ineffapp} clearly shows that such extraction of energy is only possible for inefficient measurements. In other words, efficient measurements can never yield energy~\cite{sagawaueda09,inefficient-measurements}.

	\begin{figure*}[tbh!]
		\centering
		\def\svgwidth{0.8\textwidth}
		\begingroup%
		\makeatletter%
		\providecommand\color[2][]{%
			\errmessage{(Inkscape) Color is used for the text in Inkscape, but the package 'color.sty' is not loaded}%
			\renewcommand\color[2][]{}%
		}%
		\providecommand\transparent[1]{%
			\errmessage{(Inkscape) Transparency is used (non-zero) for the text in Inkscape, but the package 'transparent.sty' is not loaded}%
			\renewcommand\transparent[1]{}%
		}%
		\providecommand\rotatebox[2]{#2}%
		\ifx\svgwidth\undefined%
		\setlength{\unitlength}{1241.10775444bp}%
		\ifx\svgscale\undefined%
		\relax%
		\else%
		\setlength{\unitlength}{\unitlength * \real{\svgscale}}%
		\fi%
		\else%
		\setlength{\unitlength}{\svgwidth}%
		\fi%
		\global\let\svgwidth\undefined%
		\global\let\svgscale\undefined%
		\makeatother%
		\begin{picture}(1,0.59756101)%
		\put(0,0){\includegraphics[width=\unitlength,page=1]{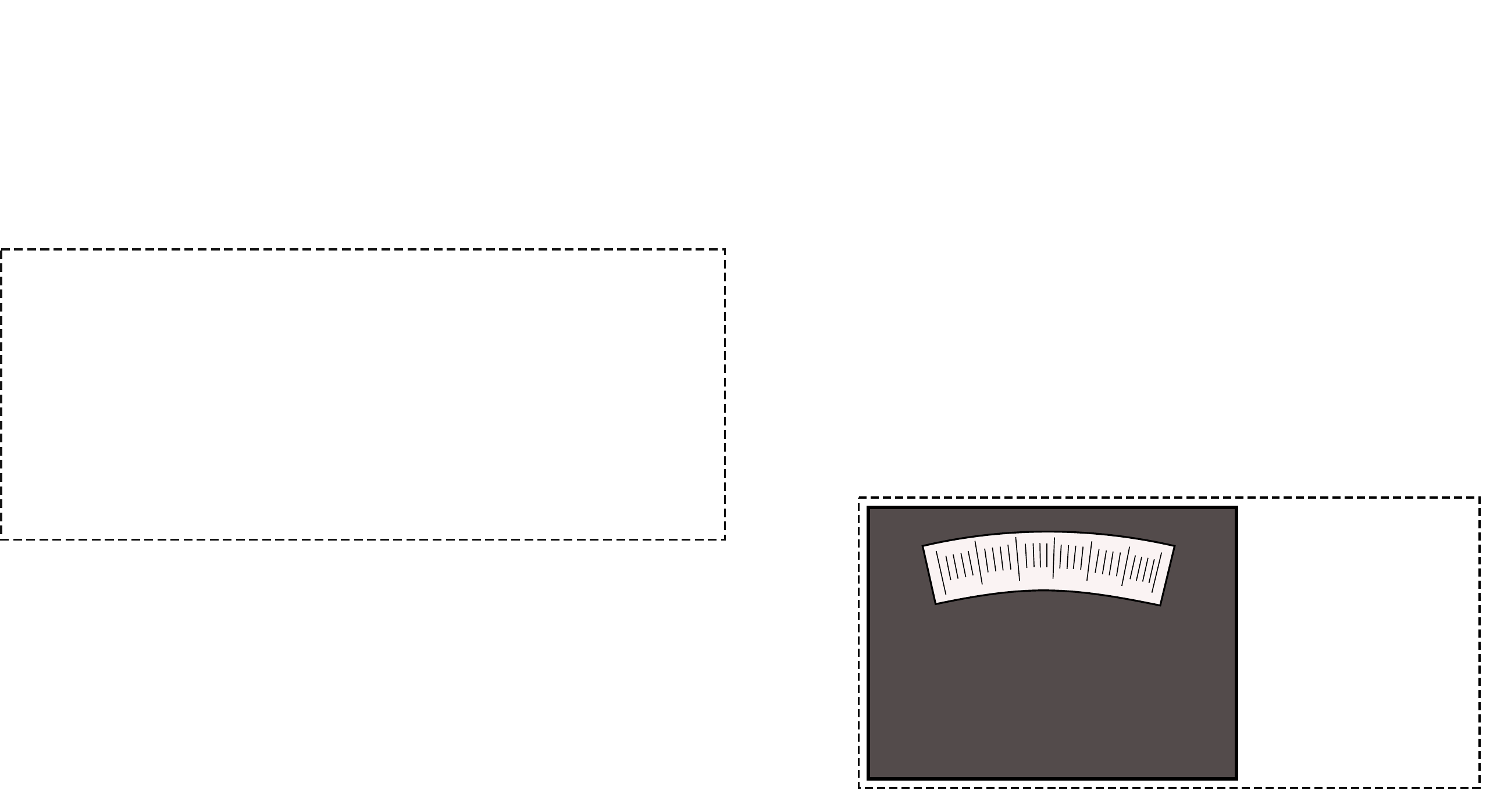}}%
		\put(0,0){\includegraphics[width=\unitlength,page=2]{fig3.pdf}}%
		\put(0,0){\includegraphics[width=\unitlength,page=3]{fig3.pdf}}%
		\put(0,0){\includegraphics[width=\unitlength,page=4]{fig3.pdf}}%
		\put(0,0){\includegraphics[width=\unitlength,page=5]{fig3.pdf}}%
		\put(0,0){\includegraphics[width=\unitlength,page=6]{fig3.pdf}}%
		\put(0.24692481,0.24083026){\color[rgb]{0,0,0}\makebox(0,0)[lb]{\white{\smash{\scalebox{0.8}[.8]{$\rho_M=|0\rangle\langle0|_{M_A} \otimes \frac{\1_{M_B}}{2}$}}}}}%
		\put(0.02234108,0.30214689){\color[rgb]{0,0,0}\makebox(0,0)[lb]{\smash{\scalebox{0.8}[.8]{$|\psi\rangle_S=c_0|0\rangle + c_1 |1\rangle$}}}}%
		\put(0.88443785,0.49768906){\color[rgb]{0,0,0}\makebox(0,0)[lb]{\smash{\scalebox{0.8}[.8]{{\bf (a)}}}}}%
		\put(0.84043785,0.46268906){\color[rgb]{0,0,0}\makebox(0,0)[lb]{\smash{\scalebox{0.8}[.8]{$\rho'_{S,k}=|k\rangle\langle k|_S$}}}}%	
	
		\put(0.60080249,0.4012209){\color[rgb]{0,0,0}\makebox(0,0)[lb]{\white{\smash{\scalebox{0.8}[.8]{$ |c_k|^2 |k\rangle \langle k|_{M_A} \otimes \frac{\1_{M_B}}{2}$}}}}}%
		
		\put(0.8844378,0.171706727){\color[rgb]{0,0,0}\makebox(0,0)[lb]{\smash{\scalebox{0.8}[.8]{{\bf (b)}}}}}%	
		\put(0.8544378,0.140706727){\color[rgb]{0,0,0}\makebox(0,0)[lb]{\smash{\scalebox{0.8}[.8]{$\rho'_{S,k}=\frac{\1_{S}}{2}$}}}}%
	
		\put(0.590194752,0.069594807){\color[rgb]{0,0,0}\makebox(0,0)[lb]{\white{\smash{\scalebox{0.75}[.8] {$|c_k|^2 |k\rangle \langle k|_{M_A} \otimes |k\rangle \langle k|_{M_B}$}}}}}%
		
		\put(0.10635392,0.506936758){\color[rgb]{0,0,0}\makebox(0,0)[lb]{\fontsize{9.5pt}{0cm}\selectfont \smash{Standard projective measurement}}}%
		\put(0.10635392,0.478804685){\color[rgb]{0,0,0}\makebox(0,0)[lb]{\fontsize{9.5pt}{0cm}\selectfont \smash{always {\bf consumes energy}:}}}%
		\put(0.1416954,0.443487013){\color[rgb]{0,0,0}\makebox(0,0)[lb]{\fontsize{9.5pt}{0cm}\selectfont \smash{$E_{\mathrm{ext}}=- k_B T H(\{p_k\})\leq 0$}}}%
		
		\put(0.10690392,0.09752820872){\color[rgb]{0,0,0}\makebox(0,0)[lb]{\fontsize{9.5pt}{0cm}\selectfont \smash{{\bf Extract energy} from}}}%
		\put(0.10690392,0.069271476){\color[rgb]{0,0,0}\makebox(0,0)[lb]{\fontsize{9.5pt}{0cm}\selectfont \smash{modified projective measurement:}}}%
		\put(0.1116954,0.0348393956){\color[rgb]{0,0,0}\makebox(0,0)[lb]{\fontsize{9.5pt}{0cm}\selectfont \smash{$E_{\mathrm{ext}}=k_B T \big( \ln2- H(\{p_k\})\big)\geq 0$}}}%
		
		\put(0.7391702948,0.27019607){\color[rgb]{0,0,0}\makebox(0,0)[lb]{\fontsize{9pt}{0cm}\selectfont \smash{swap $S$ and $M_B$}}}%
		\end{picture}%
		\endgroup%

		\caption{\textbf{Extracting energy from measurement.} A pure qubit $\rho_S=|\psi \rangle\langle \psi|_S$ is measured in the spin-$z$ basis by a device with bipartite memory $M=M_AM_B$ and initial state $\rho_M=|0\rangle\langle0|_{M_A}\otimes\ii_{M_B}/2$, in one of two different ways. Either measurement implementation yields the same outcome distribution $\{p_k\}$ and enables feedback via $Q_k=|k\rangle\langle k|_{M_A}\otimes\ii_{M_B}$ but counterintuitively, the inefficient one allows to extract useful energy $E_{\rm ext}$ from the device, in contrast to previous results \cite{sagawaueda09,kammerlanderanders}. \textbf{(a)} The measurement is projective with $M_{k}=|k\rangle\langle k|$. Operating such a device can never yield energy, $E_{\rm ext}\leq0$ in accordance with Eq.~\eqref{secondmainresulter}. \textbf{(b)} The measurement is given by $M_{ki}=|i \rangle\langle k|/\sqrt{2}$ and thus is inefficient. Intuitively, the device implements the same unitary interaction $U_{SM_AM_B}$ as in (a), but with an additional swap of the systems $S$ and $M_B$ before the projections $\{Q_k\}$. This modified projective measurement always outputs a fully mixed state $\rho'_{S,k}={\mathbbm1}_S/2$ and yields energy $E_{\rm ext}\geq0$ (see Eq.~\eqref{Eq:ze43,@}).}
		\label{workext}

	\end{figure*}

To better understand the mechanism to extract energy during the measurement process, we consider two measurements on a qubit $S$. The first one is an efficient measurement and hence, no energy can be extracted. The second is a slight variation of the first, but is inefficient and allows us to extract $k_B T\ln 2$ of energy. See Figure \ref{workext} for a brief summary.

Our first example is a rank-1 projective measurement on a qubit system $S$ with projection operators $\{|k\rangle\langle k|\}_{k=0,1}$ and we denote by $\rho_S$ and $\rho_M$ the initial state of the measured system $S$ and memory $M$, respectively. The final state of $S$ and $M$ is of the form
\begin{equation}\label{eq:finalstateexampleapp}
\rho'_{SM}=\sum_{k=0,1}p_k |k\rangle\langle k|_S\otimes \rho'_{M,k} \ ,
\end{equation}
where $p_k=\langle k|\rho_S|k\rangle$ and the states $\rho'_{M,k}$ have support on orthogonal subspaces.
% such that the outcome value $k$ is reliably stored on $M$. 
One of the measurement devices $(\rho_M,U_{SM},\{Q_k\})$ implementing this projective measurement is given as follows: let a memory $M$ consist of two qubits $M_A$ and $M_B$ with an initial state
\begin{align}
\rho_M=|0\rangle\langle 0|_{M_A} \otimes \frac{\1_{M_B}}{2} \ .
\end{align}
Additionally, we take projections $Q_k=|k\rangle\langle k|_{M_A}\otimes \1_{M_B}$ and the unitary interaction between system and memory
\begin{multline}
U_{SM}=\Big(|0\rangle\langle 0|_{S} \otimes |0\rangle\langle 0|_{M_A} \\+ |1\rangle\langle 1|_S \otimes |1\rangle\langle 0|_{M_A}
+... \Big)\otimes \1_{M_B},
\end{multline}
where the dots indicate that any unitary extension is freely chosen.
It can be directly verified that this measurement device outputs the desired final state with $\rho'_{M,k}=|k\rangle\langle k|_{M_A}\otimes \1_{M_B}/2$.

Since this first example is a projective measurement, we have already derived in Subsec.~\ref{sec:appprof} that the energy cost is 
\begin{equation}
\beta E_{\mathrm{proj}} = \beta \Delta E_S + H(\{p_k\}),
\end{equation}
implying that $E_{\mathrm{ext}}= \Delta E_S - E_{\mathrm{proj}} \leq 0$.
Hence, as explained, no energy can be extracted for any initial state of $S$.

The situation changes if we consider a slight variation, which is our second example. 
Consider the situation where we are only interested in the outcome probabilities $p_k$ of our measurement and not in the final state of $S$. We can then construct a measurement device that in addition to the previous device performs, after $U_{SM}$ but before the projections $\{Q_k\}$, a swap operation $U_{S\leftrightarrow M_B}$ between $S$ and $M_B$ (see Figure \ref{workext}). The unitary interaction between measured system and memory in this device is then given by $U_{S\leftrightarrow M_B} \circ U_{SM}$. The post-measurement state reads
\begin{equation}
\rho'_{SM}=\frac{\1_{S}}{2} \otimes \sum_{k=0,1} p_k |k\rangle\langle k|_{M_A}\otimes |k\rangle\langle k|_{M_B}.
\end{equation}
Note that the measurement device correctly outputs the outcome probabilities.
%i.e. the measurement outcomes can be read off from the memory via the projections $Q_k$ with the correct probabilities $p_k= \langle k|\rho_S|k\rangle$, 
%and therefore still allows for conditioning on the outcome. 

In contrast to the device in the first example, it always leaves the measured system in the completely mixed state. 
In this situation, We now have from Eq.~\eqref{eq:diffent} that
\begin{equation}
\beta E_{\mathrm{ext}}= \beta \Delta E_S -\beta E_{\mathrm{cost}}= \ln 2 - H(\{p_k\})  \geq 0, \label{Eq:ze43,@}
\end{equation}
implying that, if the measured system starts in any of the states $\{|k\rangle\langle k|_S\}$, this measurement device outputs $E_{\mathrm{ext}}= k_B T\ln 2$ of useful energy. 

The reason why this slight modification of the setup allows us to extract energy is that the additional swap process introduces inefficiency into the measurement: A measurement that always outputs states of the form $\rho'_{SM}=\frac{\1_{S}}{2} \otimes \rho'_M$ cannot have a one-to-one correspondence between measurement operator $M_k$ and outcome $k$. Indeed, our device implements the quantum instrument $\{T_k(\rho_S)=\sum_{i=1}^2 \frac{1}{2} |i\rangle\langle k| \rho_S |k\rangle \langle i|\}$ with inefficiency $I=2$ and hence saturates our inefficiency bound given by Ineq.~\eqref{eq:ineffapp}.

We finally comment that our energy results can be in fact considered to be statements about thermodynamic work \cite{thermo-review} as we accounted for all energetic contributions while employing unitary actions. 
It may then be surprising that, in contrast to previous findings \cite{sagawaueda09} the implementation in Fig.~\ref{workext} can extract \emph{work} from the measurement device, while still respecting the Second Law of Thermodynamics (see Appendix~\ref{App:W2nd} for the detailed discussion).

\section{Applications} \label{S:APPL}

In this section, we provide two applications of our result to fundamental tasks in quantum information processing. One is quantum Zeno stabilisation and the other is quantum error correction (QEC).
We derive the fundamental energy cost for achieving these tasks.

\subsection{Energy costs of quantum Zeno measurements}\label{label-appendix9}
Quantum Zeno stabilization~\cite{zeno-original} is a paradigmatic quantum control protocol \cite{zeno-lidar}. 
For simplicity, we here consider a one-qubit system. The task is simply to stabilise a pure state $\ket{0}$ for some time span $t$, while the system undergoes time evolution with a Hamiltonian $H_S$, which we assume to be $E\sigma_X$ with the Pauli operator $\sigma_X$ and energies $\pm E$.
Since the state $\ket{0}$ is not the eigenstate of $H_S$, the state gradually rotates to be a different state. In the stabilization protocol, we try to avoid this effect.

In quantum Zeno stabilization protocol, this task is achieved using repeated projective measurements.
More precisely, we apply the projective measurement $\{M_0=\ket{0}\bra{0},M_1=\ket{1}\bra{1}\}$, with $M_0$ the projector onto the desired state at $N$ regular time intervals $\delta t=t/N$ over the time span $t$ whilst the disturbing Hamiltonian $H_S$ is acting. 
The projective nature of the measurement makes the state indeed stabilised. This is due to the fact that the probability for the state $\ket{0}$ to be transferred to $\ket{1}$ by the time evolution scales quadratically in $\delta t$, i.e. $(\delta t)^2=(t/N)^2$. If projective measurements are applied $N$ times, the total probability for the state to be changed is given by $(t/N)^2 \times N = t^2/N$, which can be arbitrarily small if $N$ is chosen to be sufficiently large.

The question we ask here is \emph{how much energy should be invested for the quantum Zeno stabilization scheme?} On the one hand, one may naively think that the energy cost will diverge since the number $N$ of projective measurements needs to be arbitrarily large for the perfect stabilization.
On the other hand, it may not be the case because the energy cost for one projective measurement also depends on $N$. The energy cost for one measurement is in general smaller for larger $N$ since, in that case, the state to be measured is nearly close to $\ket{0}$, so that the measurement induces little effect on the state.
From these two intuitions, it will be clear that the overall energy cost is determined by the trade-off between \emph{how many times the projective measurement should be repeated to achieve a required accuracy of the stabilization?} and \emph{how much energy we should invest in each measurement?}
Here, based on the result in Subsec.~\ref{sec:appprof}, we derive the exact relation between the energy cost of the quantum Zeno stabilisation and its accuracy.

%Note that the protocol employs multiple iterations of the {\it same} projective measurement. This may be equivalently described either by considering a single measurement device that is used repeatedly or by considering multiple devices, each possibly a different implementation of that measurement. 
%However, due to our result, Eq.~\eqref{secondmainresulter} in Subsec.~\ref{sec:appprof}, which is \emph{independent} of the structure of measurement devices, it turns out that the fundamental energy cost of both approaches should be the same.

Let us denote the state on $S$ after the $n$-th measurement by $\rho_S^{(n)}=(1-\epsilon_n)|0\rangle\langle 0 | +  \epsilon_n|1\rangle\langle 1 |$. 
The probability that the process returns $|1\rangle$ after $n$ steps is then given by $\epsilon_n$. 
Hence, the fidelity~\cite{nielsenchuang} $F := \langle 0 |\rho_S^{(N)} |0\rangle$ of the state after $N$ steps is given by $1-\epsilon_N$. 
Between the measurements, the system undergoes free time evolution according to the unitary $U=\exp(-i\delta tH_S/\hbar)$ such that the probabilities after the $(n+1)$-th measurement change to $\epsilon_{n+1}=\epsilon_n \cos(E \delta t/\hbar)^2 + (1-\epsilon_n) \sin(E\delta t/\hbar)^2$. Since $\epsilon_0=0$ by assumption, it follows that
\begin{equation}
\epsilon_n= \frac{1}{2}(1-\cos (2E\delta t/\hbar)^n) =  n \left(\frac{E\delta t}{\hbar}\right)^2 + \mathcal{O}(\delta t^4).
\end{equation}
	
We now calculate the energy cost. First, the measurement at each time step projects the state to either $\ket{0}$ or $\ket{1}$, hence $\Delta E_S=0$.
From this fact and Eq.~\eqref{secondmainresulter} in Subsec.~\ref{sec:appprof}, the $n$-th measurement consumes energy $\beta E_{\mathrm{proj}}^{(n)}= H (\{\epsilon_n,1-\epsilon_n\})$. The total energy required is then given by 
\begin{align}
\beta E_{\mathrm{Zeno}}&= \sum_{n=1}^N H (\{\epsilon_n,1-\epsilon_n\}) \ .
\end{align}

We are interested in stabilisation schemes that yield high target fidelity $F$, which can be achieved by applying the measurements in shorter and shorter time scales, $\delta t=t/N \rightarrow 0$, or in other words by applying more measurements $N\rightarrow \infty$ in constant time span $t$. In this limit the higher order terms $\mathcal{O}(\delta t^4)$ of $\epsilon_n$ will not contribute to the energy cost of the measurements, so we set $\epsilon_n\simeq n \left(\frac{E\delta t}{\hbar}\right)^2$. We then have $F\simeq 1-\frac{1}{N}\left(\frac{Et}{\hbar}\right)^2$ and
\begin{widetext}
\begin{align}
\beta E_{\mathrm{Zeno}}	&= - \sum_{n=1}^N \epsilon_n \ln \epsilon_n - \sum_{n=1}^N (1-\epsilon_n) \ln (1- \epsilon_n)  \\
%\simeq &- \sum_{n=1}^N  n \left(\frac{E t}{\hbar N}\right)^2 \ln \left[ n \left(\frac{E  t}{\hbar N}\right)^2\right] - \sum_{n=1}^N \left(1- n \left(\frac{E t}{\hbar N}\right)^2\right) \ln \left[1-  n \left(\frac{E t}{\hbar N}\right)^2\right] \\
%= &- \left(\frac{E t}{\hbar}\right)^2 \left(\sum_{n=1}^N \frac{n}{N^2} \ln \frac{n}{N} + \sum_{n=1}^N \frac{n}{N^2} \ln\left[\frac{1}{N}\left(\frac{Et}{\hbar}\right)^2\right]   \right)  - \sum_{n=1}^N \left(1-n \left(\frac{E t}{\hbar N}\right)^2\right) \ln \left[1- n \left(\frac{E t}{\hbar N}\right)^2\right]\\
&\simeq - \left(\frac{E t}{\hbar}\right)^2 \left(\sum_{n=1}^N \frac{1}{N} \frac{n}{N} \ln \frac{n}{N} + \frac{N(N+1)}{2 N^2} \ln\big[1-F \big]   \right)- \sum_{n=1}^N \left(1-n \left(\frac{E t}{\hbar N}\right)^2\right) \ln \left[1- n \left(\frac{E t}{\hbar N}\right)^2\right]\\
%&= \frac{1}{2} \left(\frac{E t}{\hbar}\right)^2 \left(\frac{3}{2} - \ln[1-F]\right)	\\
&\simeq \frac{1}{2} \left(\frac{E t}{\hbar}\right)^2 \ln\left[\frac{4.5}{1-F}\right], \label{Eq;34mlp}
\end{align}
\end{widetext}
to the leading order, where we have used that $\sum_{n=1}^N \frac{1}{N} \frac{n}{N} \ln \frac{n}{N} \simeq  \int_0^1 x \ln x dx = -1/4$ and $\ln [1- n \left(\frac{E t}{\hbar N}\right)^2] \simeq - n(\frac{E t}{\hbar N})^2 $ in the limit $N\rightarrow \infty$.

Hence, we find that the total energy required for stabilisation grows logarithmically in $1/(1-F)$ for increasing target fidelity $F$ and diverges if the perfect stabilization ($F=1$) is required.
This implies that any restriction on the energy available for the stabilization scheme directly limits the achievable accuracy, which is more clear by rewriting Eq.~\eqref{Eq;34mlp} to be
\begin{equation}
F \simeq 1- 4.5 \exp \left[ - \frac{2 \hbar^2 \beta}{(E t)^2}  E_{\rm Zeno} \right]. \label{Eq:zenorrrr}
\end{equation}
Similar energy demands apply to Zeno schemes for dragging or holonomic computation \cite{zeno-lidar,zanardi-venuti} as well.

	%%%%%%%%%%%%%%%%%%%%% QEC section
	
\subsection{Energy cost of quantum error correction (QEC)}

In any quantum information processing, the system is inevitably disturbed by noise. Hence, it is of significant importance to protect information from the noise. To this end, quantum error correction (QEC) was invented~\cite{divincenzocriteria,preskill-ftqc}, where the important information is encoded onto \emph{logical} qubits ${\mathcal L}$ using a QEC code ${\mathcal C}$, consisting of a number of \emph{physical} qubits. The physical qubits are subject to noise, but logical qubits are designed to be free from noise due to the redundancy induced by the encoding.
The heart of QEC then consists in performing repeated measurements of the \emph{error syndrome} $s$ on ${\mathcal C}$ followed by suitable feedback operations $V_s$. When this error correction is performed frequently enough, reliable information processing is possible even on noisy hardware, as guaranteed by the threshold theorem \cite{preskill-ftqc}.

In this section, we consider the energy cost of quantum error correction. 
Energetic considerations are paramount in this context since syndrome measurements with feedback must be performed many times and on many qubits in a scalable set-up, which is possible only when the energy cost for one syndrome measurement is not too high.
As the error syndrome measurement makes extensive use of  projective measurements, we can exactly compute the energy cost based on our result, Eq.~\eqref{secondmainresulter}.

As a paradigmatic example, we examine the \emph{5-qubit code} ${\mathcal C}_5$~\cite{five-qubit-code}, in which the state of a single logical qubit $|\psi\rangle=\alpha_0 |0_L\rangle + \alpha_1 |1_L\rangle$ in the code space $\mathbb{C}^2_{\mathcal{L}}$, with $\alpha_0,\alpha_1\in\mathbb{C}$ and $|\alpha_0|^2+|\alpha_1|^2=1$, is encoded into the space $ \mathcal{C}_5 \equiv (\mathbb{C}^{2})^{\otimes 5}$ of five physical qubits (see e.g.~\cite{nielsenchuang} for the concrete way of encoding).
% by using the codewords below (see e.g.~\cite{nielsenchuang})
%\begin{align}
%|0_L\rangle =&\frac{1}{4}\left[|00000\rangle + |10010\rangle + |01001\rangle+ |10100\rangle \right. \notag \\
%&+ |01010\rangle+ |00101\rangle - |11011\rangle - |00110\rangle \notag \\
%&- |11000\rangle - |11101\rangle -|00011\rangle-|11110\rangle \notag \\
%&\left.-|01111\rangle-|10001\rangle - |01100\rangle - |10111\rangle \right], \notag \\
%|1_L\rangle =&\frac{1}{4}\left[|11111\rangle + |01101\rangle + |10110\rangle+ |01011\rangle \right. \notag \\
%&+ |10101\rangle+ |11010\rangle - |00100\rangle - |11001\rangle \notag \\
%&- |00111\rangle - |00010\rangle -|11100\rangle-|00001\rangle \notag \\
%&\left.-|10000\rangle-|01110\rangle - |10011\rangle - |01000\rangle \right] \notag.
%\end{align}
As an error model, we consider the amplitude damping independently acting on each physical qubit, which is given by
\begin{equation}
\mathcal{N}_{\gamma}(\rho ) = J_1 \rho J_1^{\dagger} +J_2 \rho J_2^{\dagger} 
\end{equation}	
with Kraus operators $J_1=\sqrt{\gamma} |0\rangle\langle 1|$ and $J_2=\sqrt{\1 - J_1^{\dagger} J_1}$, where $\gamma\in[0,1]$ determines the noise strength. 
Note that our formalism can be also applied to arbitrary noise models.
The noise on the physical qubits induces a certain error also on the logical qubit, changing the state to $\rho_{S,\gamma}=\mathcal{N}_{\gamma}^{\otimes 5}(|\psi\rangle \langle\psi|)$. The error can be however detected by the error syndrome measurements, described below, and can be corrected by the followed feed-back operation.

The syndrome measurements are done in the basis of each operator in $\{\mathcal{S}^1,\mathcal{S}^2,\mathcal{S}^3,\mathcal{S}^4\}$, where
\begin{align}
\mathcal{S}^1&=X\otimes Z\otimes Z\otimes X\otimes I\\
\quad \mathcal{S}^2&=I\otimes X\otimes Z\otimes Z\otimes X \notag\\
\mathcal{S}^3&=X\otimes I\otimes X\otimes Z\otimes Z\\
\mathcal{S}^4&=Z\otimes X\otimes I\otimes X\otimes Z \notag \ ,
\end{align}
where $X,Y,Z$ denote the Pauli operators and $I$ is the identity matrix. Note that all $\mathcal{S}^j$ commute and so, they are jointly measurable. 
Each syndrome measurement $\mathcal{S}^j$ has outcomes $s^j\in\{-1,1\}$ with probability $p^{(j)}_{s^j}=\tr [P^{(j)}_{s^j} \rho_{S,\gamma}]$, where $P^{(j)}_{s^j}$ denotes the projector on the subspace corresponding to the eigenvalue $s^j$.  The set of outcomes $(s^1, s^2, s^3, s^4)$ is often called the \emph{syndrome} of the error.

We here consider two implementations of the syndrome measurements. One is to use four devices, each implementing $\mathcal{S}^j$, which we call \emph{separate measurement scheme}. 
To introduce the other, we note that the same syndrome can be obtained by performing one measurement $\mathcal{S}$ with $16$ outcomes, given by projections $\{P_s\}_{s=0}^{15}$ with $P_s = P_{s^1}^{(1)} P_{s^2}^{(2)} P_{s^3}^{(3)} P_{s^4}^{(4)}$. In this case, the syndrome $s:=(s^1, s^2, s^3, s^4)$ is obtained with probability $p_s = \tr[P_s\rho_{S, \gamma}]$. We call this scheme \emph{joint measurement scheme}.
Clearly, both schemes provide the same syndrome and so, there is no difference from the viewpoint of QEC. It however does not mean they both require the same amount of energy expense. Indeed, we show in the following that the joint measurement scheme costs less energy than the separable one.

In the following, we ignore the average energy change on $S$, $\Delta E_S$, which is just for simplicity.
We find from Eq.~\eqref{secondmainresulter} that the energy cost $E_{\mathcal{C}_5}^{\mathrm{sep}}$ of the separable measurement scheme is given by
\begin{equation}
E_{\mathcal{C}^5}^{\mathrm{sep}}= k_B  T \sum_{j=1}^4 H(\{p^{(j)}_{s^j} \}_{s^j=\pm 1}) \label{eq:amrepvfd}
\end{equation}
and the cost $E_{\mathcal{C}^5}^{\mathrm{joint}}$ of the joint measurement scheme is
\begin{equation}	
E_{\mathcal{C}_5}^{\mathrm{joint}}= k_B  T H(\{p_s \}_{s=0}^{15}). \label{eq:minenergyqec}
\end{equation}
Using the relation between the Shannon entropy and the mutual information, it is straightforward to obtain
\begin{multline}
\beta (E_{\mathcal{C}^5}^{\mathrm{sep}}-E_{\mathcal{C}_5}^{\mathrm{joint}} )\\=I(S^1:S^2) +I(S^1S^2:S^3) + I(S^1S^2S^3:S^4). \label{E43qmigrep:sdf}
\end{multline}
Noting that the mutual information is non-negative, we obtain $E_{\mathcal{C}^5}^{\mathrm{joint}} \leq E_{\mathcal{C}^5}^{\mathrm{sep}}$ for any $\gamma$.
Further, the equality holds only when the measurement outcomes are uncorrelated, because the mutual informations in Eq.~\eqref{E43qmigrep:sdf} represent the correlations between measurement outcomes.
In other words, we can conclude that the joint measurement scheme can exploit correlations in the measurement outcomes to reduce the cost of the resetting step $\mathcal{R}$, and hence it costs less than the separable measurement scheme.

	\begin{figure}[tb!]
		\centering
		\def\svgwidth{0.45\textwidth} 
		\begingroup%
		\makeatletter%
		\providecommand\color[2][]{%
			\errmessage{(Inkscape) Color is used for the text in Inkscape, but the package 'color.sty' is not loaded}%
			\renewcommand\color[2][]{}%
		}%
		\providecommand\transparent[1]{%
			\errmessage{(Inkscape) Transparency is used (non-zero) for the text in Inkscape, but the package 'transparent.sty' is not loaded}%
			\renewcommand\transparent[1]{}%
		}%
		\providecommand\rotatebox[2]{#2}%
		\ifx\svgwidth\undefined%
		\setlength{\unitlength}{602.4887085bp}%
		\ifx\svgscale\undefined%
		\relax%
		\else%
		\setlength{\unitlength}{\unitlength * \real{\svgscale}}%
		\fi%
		\else%
		\setlength{\unitlength}{\svgwidth}%
		\fi%
		\global\let\svgwidth\undefined%
		\global\let\svgscale\undefined%
		\makeatother%
		\begin{picture}(1,0.58999805)%
		\put(0,0){\includegraphics[width=\unitlength,page=1]{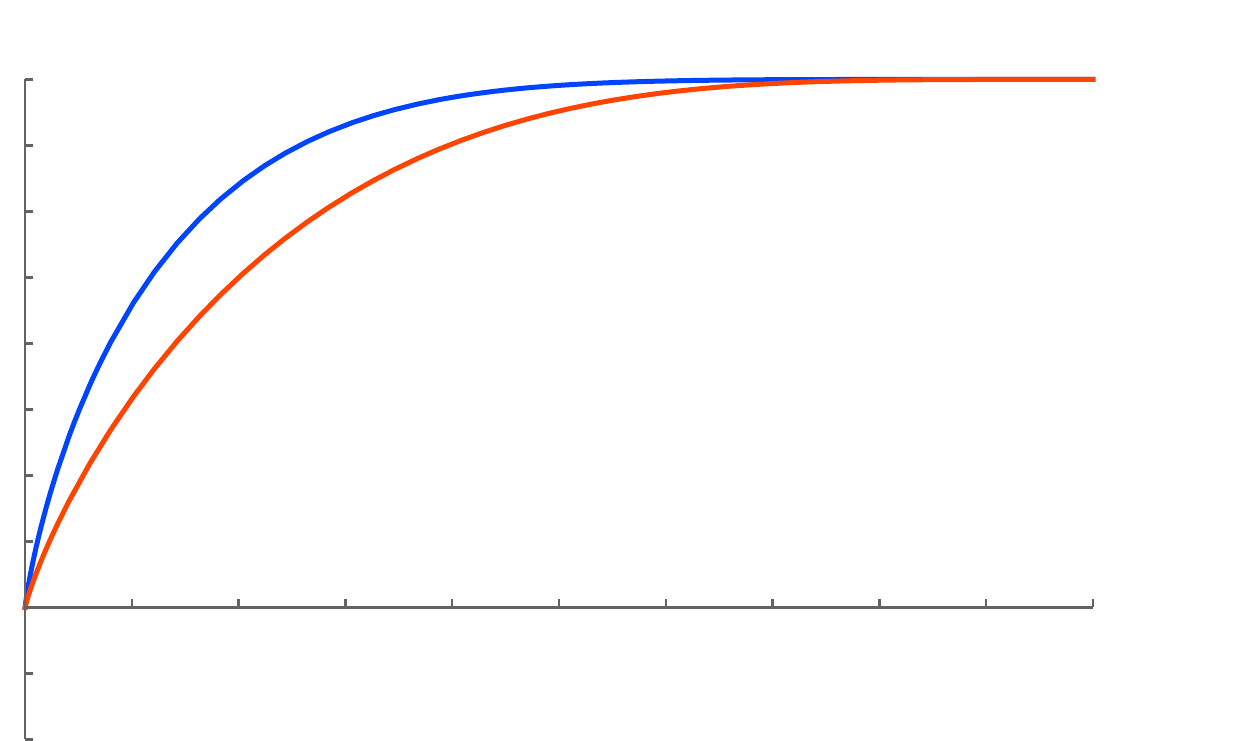}}%
		\put(-0.01612479,0.58347529){\color[rgb]{0,0,0}\makebox(0,0)[lb]{\smash{$E/(k_B T \ln2)$}}}%
		\put(0.92239114,0.08846593){\color[rgb]{0,0,0}\makebox(0,0)[lb]{\smash{$\gamma$}}}%
		\put(0.165239114,0.0601846593){\color[rgb]{0,0,0}\makebox(0,0)[lb]{\smash{$0.2$}}}%
		\put(0.335239114,0.0601846593){\color[rgb]{0,0,0}\makebox(0,0)[lb]{\smash{$0.4$}}}%
		\put(0.505239114,0.0601846593){\color[rgb]{0,0,0}\makebox(0,0)[lb]{\smash{$0.6$}}}%
		\put(0.675239114,0.0601846593){\color[rgb]{0,0,0}\makebox(0,0)[lb]{\smash{$0.8$}}}%
		\put(0.845239114,0.0601846593){\color[rgb]{0,0,0}\makebox(0,0)[lb]{\smash{$1.0$}}}%
		\put(-0.0205239114,0.09501846593){\color[rgb]{0,0,0}\makebox(0,0)[lb]{\smash{$0$}}}%
		\put(-0.045239114,-0.01){\color[rgb]{0,0,0}\makebox(0,0)[lb]{\smash{$-1$}}}%
		\put(-0.0205239114,0.2){\color[rgb]{0,0,0}\makebox(0,0)[lb]{\smash{$1$}}}%
		\put(-0.0205239114,0.304){\color[rgb]{0,0,0}\makebox(0,0)[lb]{\smash{$2$}}}%
		\put(-0.0205239114,0.408){\color[rgb]{0,0,0}\makebox(0,0)[lb]{\smash{$3$}}}%
		\put(-0.0205239114,0.516){\color[rgb]{0,0,0}\makebox(0,0)[lb]{\smash{$4$}}}%
		\put(0,0){\includegraphics[width=\unitlength,page=2]{fig4.pdf}}%
		\put(0.5500412479,0.43547529){\color[rgb]{0,0,0}\makebox(0,0)[lb]{\smash{\fontsize{7pt}{0cm}\selectfont $E/E_{\mathcal{C}_5}^{\rm joint}$}}}%
		\put(0.94239114,0.21146593){\color[rgb]{0,0,0}\makebox(0,0)[lb]{\smash{\fontsize{7pt}{0cm}\selectfont $\gamma$}}}%

		\put(0.905239114,0.17846593){\color[rgb]{0,0,0}\makebox(0,0)[lb]{\smash{\fontsize{7pt}{0cm}\selectfont $0.2$}}}%
		\put(0.7385239114,0.17846593){\color[rgb]{0,0,0}\makebox(0,0)[lb]{\smash{\fontsize{7pt}{0cm}\selectfont $0.1$}}}%
		\put(0.54385239114,0.20046593){\color[rgb]{0,0,0}\makebox(0,0)[lb]{\smash{\fontsize{7pt}{0cm}\selectfont $0.5$}}}%
		\put(0.54385239114,0.25846593){\color[rgb]{0,0,0}\makebox(0,0)[lb]{\smash{\fontsize{7pt}{0cm}\selectfont $1.0$}}}%
		\put(0.54385239114,0.313846593){\color[rgb]{0,0,0}\makebox(0,0)[lb]{\smash{\fontsize{7pt}{0cm}\selectfont $1.5$}}}%
		\put(0.54385239114,0.3713846593){\color[rgb]{0,0,0}\makebox(0,0)[lb]{\smash{\fontsize{7pt}{0cm}\selectfont $2.0$}}}%
		\end{picture}%
		\endgroup%
		
\caption{\textbf{Energy cost of quantum error correction.} For the 5-qubit code under amplitude damping noise, the cost $E_{{\mathcal C}_5}^{\rm joint}$ given by Eq.~\eqref{eq:minenergyqec} for the syndrome measurement increases with the noise level $\gamma$ {\it(red curve)}. 
When the syndrome measurement is performed separately, the cost is given by $E_{{\mathcal C}_5}^{\rm sep}$ {\it(blue)} (see Eq.~\eqref{eq:amrepvfd}).
Unlike these exact costs, only lower bounds were known such as the one $E_{{\mathcal C}_5}^{\rm SU}$ due to refs.\cite{sagawaueda09,inefficient-measurements} {\it(green)} and the naive Landauer bound \cite{landauer,nielsenLandauerQEC} $E_{{\mathcal C}_5}^{\rm Lan}$ {\it(black)}. }\label{QECfigure}
	\end{figure}

The energy cost $E_{\mathcal{C}_5}^{\mathrm{joint}}$ (red curve) and $E_{\mathcal{C}^5}^{\mathrm{sep}}$ (blue curve) are plotted in Fig.~\ref{QECfigure} as a function of the noise strength $\gamma$. 
For the sake of comparison, we also plot \emph{lower bounds} of the energy cost $E_{\mathcal{C}_5}^{\mathrm{SU}}$ of syndrome measurements obtained from the result in Ref.~\cite{sagawaueda09,inefficient-measurements} (green curve) and the cost $E_{\mathcal{C}_5}^{\mathrm{Lan}}$ from a naive application of Landauer's bound~\cite{landauer,nielsenLandauerQEC} (black curve).
Each is given by
\begin{align}
&E_{\mathcal{C}_5}^{\mathrm{SU}} = k_B T \Big[ S(\rho_{S,\gamma}) - \sum_s p_s S(P_s \rho_{S,\gamma} P_s/p_s)\Big],\\
&E_{\mathcal{C}_5}^{\mathrm{Lan}}=k_B T \big[ S(\rho_{S,\gamma}) - S(\tilde{\rho}_{S,\gamma}) \big] \leq E_{\mathcal{C}_5}^{\mathrm{SU}}.
\end{align}
We emphasise that, unlike our results $E_{\mathcal{C}_5}^{\mathrm{joint}}$ and $E_{\mathcal{C}^5}^{\mathrm{sep}}$, they are just lower bounds of the energy cost, which are not optimal in any region of $\gamma$ as seen in Fig.~\ref{QECfigure}.

	%%%%%%%%%%%%%%%%%%% CONCLUSION
	
\section{Summary and discussion}  \label{S:sd}

In this paper, based on the purely quantum-mechanical framework, we have derived a bound on the energy cost for general quantum measurement (Ineq.~\eqref{eq:firstmainresult}), consisting of the cost of the measurement itself and the cost of resetting.
%This framework is similar to the common one, but ours is slightly more general and put less assumptions, due to which we have obtained stronger results than previous ones in certain situations.
We then investigated two special cases of quantum measurement. One is a projective measurement, where we have derived the exact energy cost (Eq.~\eqref{secondmainresulter}) rather than its lower bound. 
The other is a so-called inefficient measurement, and we have shown that it is possible to extract energy in the process of measurement (Ineq.~\eqref{eq:ineffapp}) when we are only interested in the measurement outcome. By explicitly providing the setting and the measurement device which does extract a positive amount of energy, we prove that the energy extraction is indeed achievable (Eq.~\eqref{Eq:ze43,@}).

These results have been applied to two fundamental protocols in quantum information processing, quantum Zeno stabilisation and QEC. Since both are using projective measurement, we obtained their exact costs. In quantum Zeno stabilisation protocol, the energy-precision trade-off relation has been derived (Eq.~\eqref{Eq:zenorrrr}), which explicitly provides the energetic limitation of stabilising quantum states by the protocol.
On the other hand, in QEC, we have especially considered the energy cost of the so-called 5-qubit code, and have shown that the energy cost depends on how to perform the syndrome measurement. More precisely, even if two different quantum measurements offer the same syndrome, and hence result in the successful error correction, the energy cost has been shown to be less when all the syndrome is measured at once (Eqs.~\eqref{eq:amrepvfd} and~\eqref{eq:minenergyqec}). This is because the correlation between the measurement outcomes can be made use of in the resetting process.

Our result about the projective measurement (Eq.~\eqref{secondmainresulter}) also establishes a remarkable link between the viability of quantum technologies and the existence of uncertainty relations: whenever incompatible measurements have to be performed, as in quantum state tomography \cite{tomography} or quantum Monte Carlo sampling~\cite{QMCMC}, entropic uncertainty relations~ \cite{massen-uffink} yield strictly positive bounds on the entropic term, independent of the input state. Physical energy constraints together with uncertainty relations therefore place fundamental limitations on tomographic accuracy or sample quality.
	
%Our energy results are in fact statements about thermodynamic work \cite{thermo-review} as we accounted for all energetic contributions while employing unitary actions. It is then surprising that, in contrast to previous findings \cite{sagawaueda09} the implementation in Fig.\ \ref{workext} can extract useful \emph{work} from the measurement device, while still respecting the Second Law of Thermodynamics (supplementary information). Our study paves the way for investigations into the energy costs of further elementary operations\cite{bedingham-maroney} in the quantum sciences or engineering. It would be particularly interesting to extend the strength of our implementation requirement to the single-shot approach \cite{faist-renner} and quantify the arising energy fluctuations \cite{funopaper}.

Our study paves the way for investigations into the energy costs of further elementary operations~\cite{bedingham-maroney} in the quantum sciences or engineering. It would be particularly interesting to extend the strength of our implementation requirement to the single-shot approach~\cite{faist-renner} and quantify the arising energy fluctuations~\cite{funopaper}.

\section{Acknowledgements}
We thank Courtney Brell and Takahiro Sagawa for helpful comments. 
Y.N. is supported by JSPS Postdoctoral Fellowships for Research Abroad, by JSPS KAKENHI Grant Number 272650, and partially by CREST, JST, Grant No. JPMJCR1671.
D.R.\ acknowledges support from the ERC Grant DQSIM.

	%%%%%%%%%%%%%%%%%%%%%%%%%% Appendix %%%%%%%%%%%%%%%%%%%%%%%%%%%%

\onecolumn\newpage
\appendix

\section{Comparison with previous literature}\label{label-appendix8}
In this section, we compare our framework and results about the energy cost of general quantum measurement with those in the literature. In particular, we here consider those in Sagawa/Ueda~\cite{sagawaueda08,sagawaueda09}, Jacobs~\cite{inefficient-measurements}, Bennett\cite{bennettreview82,landauer}, and Faist \emph{et al.}~\cite{faist-renner}.

\subsection{Comparison with Sagawa/Ueda~\cite{sagawaueda08,sagawaueda09}}
Our framework is based on the one from Sagawa and Ueda's setting~\cite{sagawaueda08,sagawaueda09}, where they proved the following lower bound $E_{SU}$ on the energy cost for an efficient quantum measurement $\{M_k\}$:
\begin{equation}\label{eq:SagawaUeda}
E_{\mathrm{cost}}\geq E_{SU} :=  \Delta E_S + k_B T \mathcal{I},
\end{equation}
where $\mathcal{I}= S(\rho_S) + H(\{p_k\}) + \sum_k \tr [M_k \rho_S M_k \ln M_k \rho_S M_k]$. As clarified in Erratum~\cite{sagawaueda09}, the derivation relies on the fact that the measurement is \emph{efficient}. Our result given in Eq.~\eqref{eq:firstmainresult} is hence a generalisation of their result to \emph{any} measurements, including inefficient ones.
Our generalisation is proper since our bound given in Ineq.\ \eqref{eq:firstmainresult} reduces to Ineq.~\eqref{eq:SagawaUeda} (Ref.~\cite{sagawaueda09}) in the case of efficient measurements. We also note that the form of the bound given in Ref.\ \cite{sagawaueda09} does not give a correct generalization to general inefficient measurements.

\subsection{Comparison with Jacob~\cite{inefficient-measurements}}

A generalisation of the lower bound in Ineq.~\eqref{eq:SagawaUeda} to general inefficient measurements was already inferred from Jacob's work~\cite{inefficient-measurements}. 
Instead of calculating the energy expense in purely quantum mechanical set-up, the result in Ref.~\cite{inefficient-measurements} answered the converse question: \emph{after} some ``black box'' has performed the required measurement $\{M_{ki}\}$ on the state $\rho_S$, how much energy can be extracted from a feedback protocol that makes use of the measurement result $k$ and the post-measurement state $\rho'_{S,k}$? This feedback process is required to map each post-measurement state $\rho'_{S,k}$ to the initial state $\rho_S$ such that the overall process (measurement and feedback) is cyclic. It is found that the amount of energy that can be extracted in an optimal feedback process is given by the average free energy difference
\begin{equation}
E_{ext,Jacobs}= \sum_k p_k (F(\rho_S) - F(\rho'_{S,k}))=\Delta E_S + k_B T \Big[S(\rho_S) - \sum_k p_k S(\rho'_{S,k})\Big]. \label{optworkextrjacobs}
\end{equation}
By assuming the Second Law of Thermodynamics, stating that no net amount of energy can be extracted in a cyclic process that involves a single thermal bath, the same amount of energy should be invested into the measurement device, implying that
\begin{equation}
E_{cost} \geq \Delta E_S + k_B T \Big[S(\rho_S) - \sum_k p_k S(\rho'_{S,k})\Big],
\end{equation}
which is identical to our result. 

This derivation however depends on two crucial assumptions. One is that the ``measurement black-box'' acts on a thermal state (Note that in Ref.~\cite{inefficient-measurements} a non-thermal state $\rho_S$ is first reversibly transformed into a thermal state in Eq.~(7)). The other is the Second Law of Thermodynamics.
Because our results do not rely on these assumptions, they can be considered to be a \emph{quantum mechanical derivation} of the Jacob's one, or more precisely its generalisation to \emph{any} quantum states, even those far from thermodynamics equilibrium.
Our approach has strong advantages especially when we consider the cost of quantum information processing, such as those we have done in this paper, because the state during information processing is in general not a thermal state and the validity of the laws of thermodynamics is rather unclear.
Our explicit microscopic modelling is also useful e.g. in order to display the work-extracting implementation of an inefficient measurement, which is never possible in the black-box treatment in Ref.~\cite{inefficient-measurements}.

%We note however that this derivation is based on a thermodynamic framework, namely that $\rho_S$ is thermal and that the Second Law of Thermodynamics is valid, in contrast to our purely quantum-mechanical setting. Hence, this derivation is only valid for states initially in thermal equilibrium, while our first main result extends this statement to non-equilibrium states. We also note that this indirect way of computing the energy cost for measurement, based on evaluating how much energy can be extracted {\it after} the measurement and then invoking the Second Law, does not allow to prove the possibility of extracting energy {\it through} measurement in contrast to our microscopic framework (see energy extraction example).\\

\subsection{Comparison with Bennett~\cite{bennettreview82,landauer}}
We also briefly compare our results with Bennetts's \cite{bennettreview82} and other implementation-dependent bounds, such as those in Ref.~\cite{esposito-prx} for feedback protocols. 

According to Ref.~\cite{bennettreview82}, the total energy cost for measurement and erasure is $E_{\rm cost}=\Delta E_S+k_BT(S(\rho'_M)-S(\rho_M))$, where $\rho_M$ and $\rho'_M$ denote the \emph{states of the measurement device} before and after the measurement step ${\mathcal M}$. We indeed rederived this statement within our framework in Eq.\ (\ref{eq:diffent}), but argue that this result by itself is not very useful in our context, since it is \emph{not} stated in terms of the measurement specification $\{M_{ki}\}$ (and $\rho_S$, $H_S$). For any fixed measurement $\{M_{ki}\}$, one can find \emph{many} implementations $(\rho_M,U_{SM},\{Q_k\})$ yielding different values of $S(\rho'_M)-S(\rho_M)$. 
%This is well-exemplified by e.g. he red and blue curves in Fig.~\ref{QECfigure}, where the energy cost of QEC depends on the implementations that used.

All of our results are given solely in terms of the measurement specification $\{M_{ki}\}$ and the system quantities $\rho_S$, $H_S$. This feature allows us to address theoretical bounds on energy costs independently of the details of the implemented devices.

\subsection{Comparison with Faist \emph{et al.}\ \cite{faist-renner}}

In the context of general quantum operation, rather than thermodynamic operations, the energy costs for implementing quantum operation were given in Faist \emph{et al.}\ \cite{faist-renner} with a great generality.
%This work differs from all the previous more traditional thermodynamic treatments in that it goes beyond the usual ``state-transformation ideas'' and instead requires the implementation to act correctly on a larger subspace of states. 
%This is similar in spirit to our implementation requirement (Appendix \ref{measmodapp}), and 
We below comment on the two main differences between our framework and theirs. 
For simplicity, we ignore the trivial energy costs on the system $S$, i.e. setting $H_S=0$, as is done in Ref.\ \cite{faist-renner}.

In Ref.~\cite{faist-renner}, all results are derived within the so-called single-shot scenario. The energy cost is then given by the \emph{$\epsilon$-smoothed conditional max entropy}~\cite{scm, scm2}, and later translated to the asymptotic \emph{independently and identically distributed (i.i.d.)} setting.
This approach often considers the situation where the measurement device works correctly only on a certain subset of quantum states, which is controlled by the parameter $\epsilon$. 
Due to this nature of their approach, a big difference arises, as we will describe later.
Another difference arises because they make use of the post-measurement state $\rho'_S$ in the resetting step, but we do not. In the situation we are mainly interested in, $\rho'_S$ as well as the measurement outcome $k$ will be used in the later information processing, and so the post-measurement state is not available in the resetting step.
Below, we explain these differences more clearly by comparing our results with those in Ref.~\cite{faist-renner} in two most relevant cases, the zero-error single-shot case and the i.i.d. case.\\

In the zero-error single-shot case, the measurement device should act correctly for any initial state $\rho_S$, corresponding to the case where $\epsilon=0$ in  Ref.~\cite{faist-renner}. In this case, the energy cost is given by
\begin{equation}\label{eq:single-shot}
E_{\rm Faist}^{0}= k_B T  (\ln|\!|\mathcal{E}(\Pi_S)|\!|_{\infty} + \ln \mathrm{rank}[\rho'_M]),
\end{equation}
where $\Pi_S$ is the projector onto the support of the input state $\rho_S$. 
This is surely different from our result, and the difference can be intuitively understood that $E_{\rm Faist}^{0}$ is the \emph{worst-case estimate} for the energy costs in the single-shot scenario. 
It however does not mean $E_{\rm Faist}^{0}$ is always larger than or equal to our energy cost since, in their setting, the post-measurement state $\rho'_S$ is made use of in the resetting step, which saves the total energy cost.

Indeed, if we consider the situation where a projective measurement in the computational basis $\{|0\rangle,|1\rangle\}$ is performed on a pure state $(|0\rangle+|1\rangle)(\langle0|+\langle1|)/2$, then our result (Eq.~\eqref{secondmainresulter}) yields the energy cost of $E_{\rm proj}=k_BT\log2$, but $E_{\rm Faist}^{0}=0$ according to Supplementary Note 4F, Example (III) in Ref.~\cite{faist-renner}. That is, our energy cost is greater than $E_{\rm Faist}^{0}$. On the other hand, if we perform the same measurement on the state $p_0 |0\rangle_S\langle 0| + p_1 |1\rangle_S\langle 1|$, our energy cost can be computed to be $k_B T  H(\{p_0,p_1\})$, which is generally less than $E_{\rm Faist}^{0}= k_B T \log \mathrm{rank} [\{p_0,p_1\}]$. 
Again, these differences are very natural because the settings in their and our analysis are rather different.

Which framework is best suited will highly depend what one would like to know.
We think that, in the applications we have considered in this paper, i.e. quantum Zeno stabilisation and QEC, our framework is suitable since we cannot use the post-measurement state for resetting.\\

We also compare our results with the i.i.d. results in Ref.~\cite{faist-renner}, which is a little more confusing. In this case, the initial state is assumed to be $\rho_S^{\otimes n}$, where $n$ is often referred to as the number of \emph{copies}, and the asymptotic limit of $n \rightarrow \infty$ is taken in the end.
In this case, the energy cost in Ref.~\cite{faist-renner} is obtained after some simple calculation that
\begin{equation}\label{eq:iidlimit}
E_{\rm Faist}^{\rm iid}= S(\rho_S) - \sum_k p_k S(\rho'_{S,k}),
\end{equation}
which equals to our lower bound presented in Ineq.~\eqref{eq:firstmainresult} for general measurements (note that $H_S=0$), i.e. $E_{\rm Faist}^{\rm iid} \leq E_{\rm cost}$. 
From the fact that our result is just a lower bound while Eq.~\eqref{eq:iidlimit} is the exact cost, one may naively think their result is stronger than ours. This is however not the case because, again, our setting differs from theirs.
To illustrate this point more clearly, let us consider the energy cost of projective measurements, where we have also derived the exact energy cost in Eq.~\eqref{secondmainresulter}. 
Denoting our energy cost by $E_{\rm proj}$ and theirs by $E_{\rm Faist}^{\rm iid,\mathrm{proj}}$, it also follows that $E_{\rm Faist}^{\rm iid,\mathrm{proj}}\leq E_{\rm proj}$ and that the two values agree if and only if $\rho_S$ was already diagonal with respect to the measurement basis, implying that $E_{\rm Faist}^{\rm iid,\mathrm{proj}} < E_{\rm proj}$ for most initial states although both are \emph{exact} energy costs. From this example, it is clear that they are considering radically different situations.

In this case, the main difference arises from the fact that the device constructed in Ref.~\cite{faist-renner} works correctly only on a certain subspace, known as the \emph{typical subspace} of the initial state, rather than on the whole state space (see Supplementary Note 4A in Ref.~\cite{faist-renner}). The idea behind this is the information theoretic fact that the probability to obtain outcomes corresponding to the states outside the typical subspace, namely the states in the \emph{atypical subspace}, becomes arbitrarily small if the number $n$ of copies is sufficiently large. Hence, even if the device does not work correctly on the atypical subspace, the error can be ignored in the asymptotic limit, $n \rightarrow \infty$.
%When $n$ is large but finite, the error originated from ignoring the atypical subspace remains, which can be evaluated by the smoothed max entropy.
This situation is rather different from ours, and we require something \emph{stronger}. That is, we require that the measurement device should act correctly on the \emph{full} state space. Due to this stronger requirement, our energy cost result exceeds the one in Ref.~\cite{faist-renner}.

Whether to use the i.i.d. limit of Ref.~\cite{faist-renner} or our results again depends on the physical situation. We however argue that in many applications to quantum information processing, including those we have studied in this paper, our setting is more suitable since the state is in general not in the form of  $\rho^{\otimes n}$, rather it is an entangled state  $\rho_{\rm ent}^{(n)}$.

\section{Energy cost of the dephasing operation in the measurement step $\mathcal{M}$}\label{sec:projectionsokay}

In the main text, we claimed that the projections $\{Q_k\}$ onto the different subspaces $\mathcal{H}_k$ of the memory $M$ employed during the measurement step $\mathcal{M}$ is effectively seen as a dephasing operation, and that the dephasing operation can be implemented unitarily without any energy costs using an environmental system $E$. Here we prove this statement. 
This result may be of independent interest~\cite{bedingham-maroney}. 
We also point out that the following result actually improves the main result of Ref.~\cite{kammerlanderanders}: the lower bound from Ref.~\cite{kammerlanderanders} on the energy cost of a dephasing operation is negative whenever it changes the state, whereas we will show that it cab be always exactly zero.

Let $T_M(\sigma_M)=\sum_k Q_k \sigma_M Q_k$ be the dephasing operation on the memory $M$. Then, for any Stinespring dilation $U_{ME}$~\cite{Sti,wolfscript} of $T_M$, i.e.
\begin{equation}\label{eq:unitaryimplementation}
T_M(\sigma_M)=\tr_E[U_{ME}(\sigma_M\otimes\sigma_E)U_{ME}^{\dagger}], \qquad \forall \sigma_M,
\end{equation}
where $\sigma_E$ is a thermal state on an environment $E$ with Hamiltonian $H_E$, we define the corresponding energy cost by
\begin{equation}
E_{\mathrm{deph}}:= \tr  [H_{ME}  (U_{ME}(\sigma_M\otimes\sigma_E)U_{ME}^{\dagger}-\sigma_{M}\otimes \sigma_E)].
\end{equation}
Recalling that the Hamiltonian of the memory is given by $H_M=\bigoplus_{k=1}^K H_k$, where $H_k$ is a Hamiltonian on the respective subspace $\mathcal{H}_k$ with corresponding projection $Q_k$, $[Q_k,H_k]=0$, which implies that the dephasing operation  $T_M$ does not change the average energy on $M$. Hence, all energy expenses of this implementation are due to energy changes in the environment $E$, i.e.
\begin{equation}\label{eq:energyCostProjEnv}
E_{\mathrm{deph}}=\tr [H_{E} (\sigma'_{E}-\sigma_{E})].
\end{equation}

The initial state of the environment is, by assumption, thermal, i.e. $\sigma_E=\exp(-\beta H_E)/Z_E$ with $Z_E$ the partition function. The final state of $E$ on the other hand can be characterised by using the statement in Appendix~\ref{App:Lemma}, from which we obtain that there exist states $\sigma_{E,k}$ with $S(\sigma_{E,k})=S(\sigma_{E})$ for all $k$ such that
\begin{equation}\label{eq:proofappenerg}
\sigma_E'= \sum_k \tr [Q_k \sigma_M] \sigma_{E,k}, \qquad \forall \sigma_M.
\end{equation}
Substituting this, and using the fact that thermal state minimizes the average energy on an entropic orbit, $\beta \tr [H_{E} (\sigma_{E,k}-\sigma_{E})]= D(\sigma_{E,k},\sigma_E) \geq 0$ for all $k$, where $D(\rho,\sigma)$ is the relative entropy and is non-negative, it follows that
\begin{align}
E_{\mathrm{deph}}&=\tr [H_{E} (\sigma'_{E}-\sigma_{E})] =\sum_k  \tr[Q_k \sigma_M]  \tr [H_{E} (\sigma_{E,k}-\sigma_{E})] \geq 0. \label{eq:vanishcost}
\end{align}		
That is, the energy cost of implementing the dephasing operation $T_M$ is always non-negative.

We further show that there exist $\sigma_E$ and $U_{ME}$ such that the energy cost $E_{\mathrm{deph}}$ of the corresponding dilation of $T_M(\sigma_M)$ is precisely zero. To this end, we first provide a characterisation of all unitaries $U_{ME}$ which, given any fixed full rank state $\sigma_E$, satisfy Eq.~\eqref{eq:unitaryimplementation}, or more precisely,
\begin{equation}\label{eq:projrequ} 
\tr_E[U_{ME}(\sigma_M\otimes\sigma_E)U_{ME}^{\dagger}]= \sum_k Q_k \sigma_M Q_k, \qquad \forall \sigma_M.
\end{equation}
In the end, we will describe a simple explicit construction.
	
Consider a pure state $\sigma_M=\psi=|\psi\rangle \langle \psi|$ on $M$ in the support of a fixed projector $Q_k$, i.e. $Q_k \psi=\psi$. From Eq.~\eqref{eq:projrequ}, the marginal on $M$,
\begin{equation}
\tr_E[U_{ME}(\psi\otimes \sigma_E)U_{ME}^{\dagger}]=\psi,
\end{equation}
is pure, while by Eq.~\eqref{eq:proofappenerg}, the marginal state $\sigma_E'$ on $E$ is
\begin{equation}
\tr_S[U_{ME}(\psi\otimes \sigma_E)U_{ME}^{\dagger}] = \sigma_{E,k},
\end{equation}
where $\sigma_{E,k}$ is a state which is independent of $\psi$. This implies that
\begin{equation}\label{eq:projunit1}
U_{ME} (\psi \otimes \sigma_E) U_{ME}^{\dagger}= \psi\otimes \sigma_{E,k}.
\end{equation}
Let us now denote the spectral decomposition of $\sigma_E$ by $\sigma_E=\sum_j \lambda_j |\phi_j\rangle\langle \phi_j|_E$. We note that all eigenvalues $\lambda_j$ are strictly positive since $\sigma_E$ is assumed to have full rank. Substituting the decomposition of $\sigma_E$ into Eq.~\eqref{eq:projunit1} we obtain
\begin{equation}\label{eq:projunit2}
\sum_j \lambda_j U_{ME} (\psi \otimes |\phi_j\rangle\langle \phi_j|_E) U_{ME}^{\dagger}= \psi\otimes \sigma_{E,k}.
\end{equation}
We therefore obtain that 
\begin{equation}
\tr_E \Big[U_{ME} (\psi \otimes |\phi_j\rangle\langle \phi_j|_E) U_{ME}^{\dagger}\Big] = \psi, \qquad \forall j,
\end{equation} 
which further implies that the unitary $U_{ME}$ that satisfies Eq.~\eqref{eq:projrequ} with full rank states $\sigma_E$ is of the form
\begin{equation}\label{eq:unitaryForProj}
U_{ME}\left(|\psi\rangle_M \otimes |\phi_j\rangle_E \right)=|\psi\rangle_M \otimes V_{\psi}|\phi_j\rangle_E, 
\end{equation}
with a unitary $V_{\psi}$ depending on $\psi\in Q_k$. Note however that the right-hand side of Eq~\eqref{eq:unitaryForProj} must be linear in $\psi$ due to the linearity of the left-hand side. Hence, $V_{\psi}$ can only depend on the label $k$ of the subspace corresponding to $Q_k$, so we write $V_k=V_{\psi}$.
	
To characterise the unitaries $V_k$, we next evaluate Eq.~\eqref{eq:projrequ} on a mixed initial state $\sigma_M$ on $M$.
From Eq.~\eqref{eq:unitaryForProj}, it follows that
\begin{align}
U_{ME}(\sigma_M\otimes \sigma_E)U_{ME}^{\dagger}= U_{ME}\Big(\sum_{ij}Q_{i}\sigma_M Q_j \otimes \sigma_E\Big)U_{ME}^{\dagger} 	= \sum_{ij}Q_{i}\sigma_M Q_j\otimes V_i\sigma_E V_j^{\dagger}.
\end{align}
which together with Eq.~\eqref{eq:projrequ} leads to 
\begin{equation}
\sum_{ij} \tr _E \big[V_i\sigma_E V_j^{\dagger}\big] Q_{i}\sigma_M Q_j  =\sum_k Q_k  \sigma_M Q_k, \qquad \forall \sigma_M.
\end{equation}
This holds if and only if the unitaries $V_k$ satisfy
\begin{equation}\label{eq:newScalarProduct}
\tr [V_i\sigma_E V_j^{\dagger} ]=\delta_{ij}, \quad \forall i,j,
\end{equation}	
implying that the unitaries $V_k$ must form an orthonormal unitary operator basis with respect to the modified scalar product Eq.~\eqref{eq:newScalarProduct}. Such orthonormal unitary operator bases only exist if the Hilbert space dimension $d_E$ of the environmental system $E$ is sufficiently large compared to the number $K$ of possible outcomes $k$, i.e. $d_E\geq \sqrt{K}$. For further properties of unitary operator bases we refer to Ref.~\cite{wolfscript}.
Thus, given a full rank state $\sigma_E$ on $E$, any unitary $U_{ME}$ satisfying Eq.~\eqref{eq:projrequ} is of the form Eq.~\eqref{eq:unitaryForProj} with unitaries $V_k$ that meet the condition Eq.~\eqref{eq:newScalarProduct}.

To show that there exists an implementation of the dephasing channel \eqref{eq:projrequ} with vanishing energy cost $E_{\mathrm{deph}}$, we may therefore choose the Hamiltonian of the environment $E$ to be trivial, $H_E=0$, implying that the initially thermal state of $E$ is maximally mixed for all $\beta$,  $\sigma_E=\1_E/d_E$, and that the energy cost is $E_{\mathrm{deph}}=0$ by Eq.~\eqref{eq:energyCostProjEnv}. The corresponding unitary $U_{ME}$ that implements the dephasing channel is given by $U_{ME}=\sum_k Q_k \otimes V_k$ with unitary operators $V_k$ satisfying $\tr [V_i V_j^{\dagger} ]= d_E\delta_{ij}$ for all $i,j$, which can be easily checked that
\begin{align}
\tr_E[U_{ME}(\sigma_M\otimes \sigma_E)U_{ME}^{\dagger}]%&= \tr_E\Big[\Big(\sum_j Q_j\otimes V_j\Big) \Big(\sigma_M\otimes \1_E/d_E\Big) \Big(\sum_k Q_k \otimes V_k^{\dagger}\Big)\Big]\\
%&=\frac{1}{d_E} \sum_{j,k}  Q_j \sigma_M Q_k \tr[V_j V_k^{\dagger}]\\
&=\sum_k Q_k \sigma_M Q_k.
\end{align}
More concretely, the unitaries $V_k$ can for example be chosen as distinct elements from the set of unitaries
\begin{equation}
V_{l,m}=\sum_{r=0}^{d_E-1} \mathrm{e}^{\frac{2\pi i}{d_E} r m}|l+r\rangle\langle r|, \qquad l,m=0,1,...,d_E-1,
\end{equation}
where the addition in $|l+r\rangle$ is taken modulo $d_E$. These $d_E^2$ operators can be understood as a discrete version of the Heisenberg-Weyl operators and indeed satisfy, as one can easily compute,
\begin{equation}
\tr [V_{l,m} V_{s,t}^{\dagger}]= d_E \delta_{l,s} \delta_{m,t}, \qquad \forall l,m,s,t\in\{0,...,d_E-1\}.
\end{equation}

\section{The states after projective measurements} \label{App:Lemma}

We here show an important statement about the states after projective measurements, which are used in Sebsec.~\ref{sec:appprof} and Appendix~\ref{sec:projectionsokay}. More precisely, we show the following. 
Let $\{P_k\}$ be a projective measurement on a quantum system $S$. Let the Stinespring dilation of the ``measurement channel'' $T_S(\rho_S):=\sum_k P_k \rho_S P_k$ be given by
\begin{equation}\label{eq:projectionrequirement}
\sum_k P_k \rho_S P_k = \tr_E\left[U_{SE} (\rho_S\otimes \rho_E) U_{SE}^{\dagger}\right], \quad \forall \rho_S,
\end{equation}
where $\rho_E$ is an initial state of a quantum system $E$ and $U_{SE}$ is a unitary on $S$ and $E$.
Then, there exist quantum states $\sigma_{E,k}$ with $S(\rho_E) = S(\sigma_{E,k})$ for all $k$ such that the post-measurement state of $E$, $\rho_E'=\tr_S\left[ U_{SE} (\rho_S\otimes \rho_E) U_{SE}^{\dagger}\right]$, can be written as 
\begin{equation}
\rho_E'= \sum_k \tr [P_k \rho_S] \sigma_{E,k}, \quad \forall \rho_S.\label{lem:form}
\end{equation}
If, additionally, $U_{SE}$ and $\rho_E$ together with projections $\{Q_k\}$ on $E$ form an implementation $(U_{SE},\rho_E, \{Q_k\})$ of the projective measurement $\{P_k\}$ on $S$, and 
\begin{equation}\label{additionalapp}
P_k \rho_S P_k = \tr_E\left[(\1 \otimes Q_k) U_{SE} (\rho_S\otimes \rho_E) U_{SE}^{\dagger} (\1 \otimes Q_k)\right], \quad \forall \rho_S,\  \forall k,
\end{equation}
then $\sigma_{E,k}=Q_k \sigma_{E,k}Q_k$ for all $k$, i.e. the $\sigma_{E,k}$ are mutually orthogonal.

The proof is based on the Stinespring dilation theorem, according to which we can always write the channel $T_S(\rho_S)$ as a unitary $U_{SA}$ acting on $S$ and an ancilla $A$ initially in a pure state $|0\rangle\langle 0 |_A$ such as
\begin{equation}
T_S(\rho_S)=\tr_A[U_{SA}(\rho_S \otimes |0\rangle\langle 0 |_A) U_{SA}^{\dagger}].
\end{equation}
The minimal Stinespring dilation can be chosen to be any unitary extension $U_{SA}$ of the operator $\sum_k P_k \otimes |k \rangle \langle 0 |_A$, whose action is only defined on states of the form $|\psi\rangle_S\otimes |0\rangle_A$, where the ancilla Hilbert space $A$ is spanned by the orthonormal basis $|k\rangle$~\cite{nielsenchuang}. The corresponding complementary channel takes the form
\begin{align}\label{eq:onecompchannel}
T_{A}(\rho_S) &:= \tr_S[U_{SA}(\rho_S \otimes |0\rangle\langle 0 |_A) U_{SA}^{\dagger}]  \notag \\
%	&=\tr_S \Big[\sum_{k,k'} P_k \rho_S P_{k'} \otimes |k\rangle_A\langle k' |\Big] \ , \notag \\
&=\sum_{k} \tr[P_k \rho_S ]  |k\rangle\langle k |_A.
\end{align}
This channel is not the only possible complementary channel of $T_S$. Using Eq.~\eqref{eq:projectionrequirement}, we find another complementary channel, 
\begin{equation}
T_{E\tilde{E}}(\rho_S)=\tr_S [U_{SE\tilde{E}} (\rho_S \otimes \psi_{E\tilde{E}})U_{SE\tilde{E}}^{\dagger}] \ ,
\end{equation}
where $\tilde{E}$ is a purifying system of $E$ such that the pure state $\psi_{E\tilde{E}}$ satisfies $\tr_{\tilde{E}}[\psi_{E\tilde{E}}]=\rho_E$ and $U_{SE\tilde{E}}:=U_{SE}\otimes \1_{\tilde{E}}$. The Stinespring theorem states that these two complementary channels, $T_{A}$ and $T_{E\tilde{E}}$ are related by an isometry $V: \mathcal{H}_A\rightarrow\mathcal{H}_E \otimes \mathcal{H}_{\tilde{E}}$. Hence, we obtain
\begin{equation}
T_{E\tilde{E}}(\rho_S)=\sum_{k} \tr[P_k \rho_S ]  |\gamma_k\rangle\langle \gamma_k |_{E\tilde{E}}
\end{equation} 
with $|\gamma_k \rangle_{E\tilde{E}}:= V |k \rangle_A$ forming an orthonormal basis. 
Note that the complementary channel $T_{E\tilde{E}}$ and the final state $\rho'_E$ of $E$ are, by construction, linked via the partial trace,
\begin{align}
\rho'_E&=\tr_S \left[U_{SE} (\rho_S\otimes \rho_E) U_{SE}^{\dagger}\right] = \tr_{\tilde{E}} \left[T_{E\tilde{E}}(\rho_S)\right] \ .
\end{align}
Hence, the final state on $E$ takes the form
\begin{equation}\label{rhoEPrimeDecomp}
\rho'_E = \sum_k \tr[P_k \rho_S ] \sigma_{E,k}, \qquad \forall \rho_S,
\end{equation}
where we define the states $\sigma_{E,k}=\tr_{\tilde{E}} [V |k\rangle\langle k |_AV^{\dagger}]$, which are independent of $\rho_S$. 
	
To show $S(\rho_E)=S(\sigma_{E,k})$, let now $\rho_S=\psi_k$ be a pure state supported on the subspace characterised by one $P_k$, i.e. $P_k \psi_k=\psi_k$. Then by Eq.~\eqref{rhoEPrimeDecomp}
\begin{equation}\label{addtionalapp2}
\rho'_E = \tr_S[U_{SE}(\psi_k\otimes \rho_E)U_{SE}^{\dagger}] = \sigma_{E,k}, 
\end{equation}
and the final state on $S$ is pure,
\begin{equation}\label{eq:finalStateOnE}
\rho_S'=\tr_E[U_{SE}(\psi_k\otimes \rho_E)U_{SE}^{\dagger}]=\sum_{k'} P_{k'} \psi_k P_{k'} =\psi_k.
\end{equation}
Hence there are no correlations between the marginals of the final $SE$ state, i.e. $U_{SE}(\psi_k\otimes \rho_E)U_{SE}^{\dagger}=\psi_k\otimes \sigma_{E,k}$. Since unitaries do not change the spectrum, we have $S(\rho_E)=S(\sigma_{E,k})$, which concludes the first part of the proof.%, which shows that $\sigma'_{E,k}= W_k \sigma_E W_k^{\dagger}$ where $W_k$ are unitaries.\\
	\\
	
For the second part of the proof, we start with Eq.~\eqref{additionalapp}. By summing Eq.~\eqref{additionalapp} over all $k$, we obtain Eq.~\eqref{eq:projectionrequirement}, implying that all statements within the first part of the proof remain valid. We can thus take $\psi_k$ to be a state in the support of $P_k$ as above to find by Eq.~\eqref{addtionalapp2} that
\begin{align}
\sigma_{E,k}&=\tr_S\big[U_{SE}(\psi_k\otimes \rho_E)U_{SE}^{\dagger}\big].
\end{align}
%	Our aim is to show that by requiring \eqref{additionalapp} we have, for all $k$, that
%	\begin{equation}\label{addapp3}
%	\sigma_{E,k}=Q_k \sigma_{E,k} Q_k \ . 
%	\end{equation}
We then observe that the quantity $Q_k \sigma_{E,k} Q_k$ is a positive operator with unit trace for all $k$ since it follows from Eq.~\eqref{additionalapp} that
\begin{align}
\tr_E \big[Q_k \sigma_{E,k}Q_k \big] &=  \tr  _{SE}\big[(\1 \otimes Q_k)U_{SE}(\psi_k\otimes \rho_E)U_{SE}^{\dagger}(\1 \otimes Q_k)\big] \notag \\
&= \tr_S\big[P_k\psi_k P_k\big] = \tr \psi_k =1.
\end{align}
Thus, we have
\begin{align}
1= \tr [\sigma_{E,k}] &= \tr [(Q_k + (\1-Q_k))\sigma_{E,k}(Q_k + (\1-Q_k))] \notag \\
&= \tr [Q_k \sigma_{E,k} Q_k ] + \tr [ (\1-Q_k) \sigma_{E,k}  (\1-Q_k) ] \notag \\
&= 1  + \tr [ (\1-Q_k) \sigma_{E,k}  (\1-Q_k) ],
\end{align}
which implies that $(\1-Q_k) \sigma_{E,k} =\sigma_{E,k}(\1-Q_k) =0$. Thus, we obtain $\sigma_{E,k}=Q_k \sigma_{E,k} Q_k$ for all $k$.

\section{Work cost of quantum measurements and the Second Law of Thermodynamics} \label{App:W2nd}
In this section we argue why our results of energy costs developed in this paper are, in fact, also statements about thermodynamic work.
More concretely, we argue that, by identifying the projective measurements with the dephasing operations and using the result in Appendix~\ref{sec:projectionsokay}, all energy costs of an implementation of a quantum measurement can be considered to stem from unitary dynamics $U$. Hence, thermodynamic work cost is given as the average energy change \cite{Karenworkdef,Andersworkdef,thermo-review}
\begin{equation}
\label{eq:defwork}
W= \tr [H (U \rho U^{\dagger} - \rho)] \ 
\end{equation}
of a system with Hamiltonian $H$ initially in the state $\rho$, which is equal to the energy cost $E_{\mathrm{cost}}$ in our results.
%Indeed, the energy expenses of the measurement step $\mathcal{M}$ are due to the unitary $U_{SM}$ and not the projections $\{Q_k\}$ as shown in Appendix \ref{sec:projectionsokay} and the resetting step is unitary by construction. 
%Hence, the overall work cost $W_{\mathrm{cost}}$ of conducting a general quantum measurement is exactly equal to the 

As a consequence of this identification of our energy cost and work, the energy extraction example in Subsec.~\ref{sec:appinefficiency} illustrates that  
\emph{extracting useful thermodynamic work from the measurement device is possible}. This may seem intriguing in the context of the Second Law of Thermodynamics: discussions on the net work gain in a whole cycle of a Szilard engine typically assume that no work can be extracted in the measurement itself and argue that all work gained in the extraction phase of the cycle is completely cancelled by the cost imposed by Landauer's principle~\cite{landauer} for resetting the memory that stores the measurement outcome~\cite{SzilardMeasNoCost}.

The work extraction from the measurement process however does not contradict the Second Law of Thermodynamics for the following reason. When considering the overall work gain of a Szilard engine that employs a measurement device as described in our work extraction example in Subsec.~\ref{sec:appinefficiency}, one needs to incorporate the cost of completing the thermodynamic cycle by restoring the initial pure state on $S$. This restoring step consumes all work gained during the measurement. Indeed, one finds from Eq.~\eqref{eq:firstmainresult} that
\begin{align}
W_{\mathrm{cost}}&\geq \Delta E_S + k_B T \Delta S \\
&= \Delta F_S + k_B T (S(\rho'_S) -S(\rho_S)) + k_B T \Delta S \\
&= \Delta F_S + k_B T \Big(S(\rho'_S) - \sum_k p_k S(\rho'_{S,k})\Big)\\
&\geq \Delta F_S, \label{eq:2ndlaw}
\end{align}
where $\Delta F_S = F_S(\rho'_S) - F_S(\rho_S)$ is the free energy change in the system during the measurement, which corresponds to the work cost of the aforementioned restoring step of the measured system $S$. 
%The inequality \eqref{eq:2ndlaw} follows from the concavity of the von Neumann entropy.
This shows that the overall work expense in a full thermodynamic cycle that includes measurements is always non-negative and shows the validity of the Second Law in our general setting, as expected.

%\bibliography{scibib}

%\bibliographystyle{Science}

\end{document}